\documentclass[pdflatex,sn-mathphys-num]{sn-jnl}% Math and Physical Sciences Numbered Reference Style
\usepackage{graphicx}%
\usepackage{multirow}%
\usepackage{amsmath,amssymb,amsfonts}%
\usepackage{amsthm}%
\usepackage{mathrsfs}%
\usepackage[title]{appendix}%
\usepackage{xcolor}%
\usepackage{textcomp}%
\usepackage{manyfoot}%
\usepackage{booktabs}%
\usepackage{algorithm}%
\usepackage{algorithmicx}%
\usepackage{algpseudocode}%
\usepackage{listings}%
\usepackage[utf8]{inputenc}

\usepackage[ngerman,american]{babel}

\usepackage{microtype}

\usepackage{times}

\usepackage{amsmath,bm}

\usepackage{graphicx}

\usepackage{xspace}

\usepackage{units}

\usepackage{color}

\usepackage{amsthm}
\newtheorem{remark}{Remark}

\usepackage{booktabs}
\usepackage{longtable}
\usepackage{tabularx}
\usepackage{array}

\usepackage{subfigure}

\usepackage{hyperref}

\usepackage{pdflscape}
\usepackage[figuresright]{rotating}

\usepackage{cleveref}
\crefrangeformat{equation}{(#3#1#4) to~(#5#2#6)}
\crefmultiformat{equation}{(#2#1#3)}{ and~(#2#1#3)}{, (#2#1#3)}{ and~(#2#1#3)}

\usepackage[numbers,sort&compress]{natbib}

\usepackage{xfp}

\usepackage{enumerate} 

\usepackage{tikz}
\usepackage{pgfplots}
\pgfplotsset{compat=1.16}
\usetikzlibrary{arrows.meta}
\pgfplotsset{
    every axis/.append style={
        width=5cm,
        height=3.09016994cm
    }
}

\makeatletter
\renewcommand\@pnumwidth{2em}%
\makeatother

\usepackage{ifdraft}

\definecolor{mygreen}{rgb}{0.0,0.63,0.0}

\usepackage{etoolbox}
\usepackage{xifthen}
\usepackage{bm}
\usepackage{amssymb}
\usepackage{mathtools}

\newcommand{\eg}{e.g.\xspace}
\newcommand{\cf}{cf.~}

\newcommand{\sr}{Simo--Reissner\xspace}
\newcommand{\kl}{Kirch\-hoff--Love\xspace}

\newcommand{\R}[1]{\mathbb{R}^{#1}}
\newcommand{\SO}{SO(3)}

\newcommand{\order}[1]{\mathcal O(#1)}
\newcommand{\e}[1]{\tns{e}_{#1}}

\newcommand{\ex}{\e{1}}
\newcommand{\ey}{\e{2}}
\newcommand{\ez}{\e{3}}

\newcommand{\pfrac}[2]{\frac{\partial #1}{\partial #2}}

\newcommand{\tn}[2]{%
\ifnumcomp{#1}{=}{1}{\underline{\boldsymbol{#2}}}% First order tensor
{\ifnumcomp{#1}{=}{2}{\underline{\boldsymbol{#2}}}% Second order tensor
{\ifnumcomp{#1}{>}{2}{Higher order tensor not yet implemented!}% Higher order tensors
{Wrong tensor order given}%
}}}
\newcommand{\tns}[1]{\tn{1}{#1}}
\newcommand{\tnss}[1]{\tn{2}{#1}}
\newcommand{\tnsO}{\tns{0}} % Zero first order tensor
\newcommand{\tnssI}{\tnss{I}} % Identity second order tensor
\newcommand{\vv}[1]{\boldsymbol{\mathsf{#1}}}
\newcommand{\mat}[1]{\boldsymbol{\mathsf{#1}}}

\newcommand{\matI}{\mat{I}}
\newcommand{\norm}[1]{\left\|#1\right\|}
\newcommand{\tr}{^{\mathrm T}}
\renewcommand{\vector}[1]{\begin{bmatrix}#1\end{bmatrix}}
\renewcommand{\matrix}[1]{\begin{bmatrix}#1\end{bmatrix}}

\newcommand{\brackets}[4][]{%
\ifthenelse{\isempty{#1}}{%
\left#2#4\right#3%
}{%
\ifnumcomp{#1}{=}{0}{#2#4#3}% Default size
{\ifnumcomp{#1}{=}{1}{\bigl#2#4\bigr#3}% bigl
{\ifnumcomp{#1}{=}{2}{\Bigl#2#4\Bigr#3}% Bigl
{\ifnumcomp{#1}{=}{3}{\biggl#2#4\biggr#3}% biggl
{\ifnumcomp{#1}{=}{4}{\Biggl#2#4\Biggr#3}% Biggl
{size not supported}
}}}}}}

\newcommand{\br}[2][]{\brackets[#1]{(}{)}{#2}}
\newcommand{\equationAndInline}[2]{\mathchoice{#1}{#2}{#1}{#1}}

\newcommand{\placeholder}{(\cdot)}

\newcommand{\Ebeam}{E}
\newcommand{\EbeamAvg}{\bar{E}}
\newcommand{\EbeamA}{E_1}
\newcommand{\EbeamB}{E_2}

\newcommand{\nubeam}{\nu}

\DeclareMathOperator{\rv}{rv}

\newcommand{\triad}{\tnss{\Lambda}}
\newcommand{\triadi}{\tnss{\Lambda}_i}
\newcommand{\triadA}{\tnss{\Lambda}_1}
\newcommand{\triadB}{\tnss{\Lambda}_2}
\newcommand{\triadBA}{\tnss{\Lambda}_{21}}

\newcommand{\triadiO}{\tnss{\Lambda}^0_i}
\newcommand{\triadAO}{\tnss{\Lambda}^0_1}
\newcommand{\triadBO}{\tnss{\Lambda}^0_2}

\newcommand{\triadTildei}{\hat{\tnss{\Lambda}}_i}
\newcommand{\triadTildeA}{\hat{\tnss{\Lambda}}_1}
\newcommand{\triadTildeB}{\hat{\tnss{\Lambda}}_2}
\newcommand{\triadTildeBA}{\hat{\tnss{\Lambda}}_{21}}
\newcommand{\triadTildeAB}{\hat{\tnss{\Lambda}}_{12}}
\newcommand{\rotvecTildeBA}{\hat{\tns{\psi}}_{21}}
\newcommand{\rotvecTildeAB}{\hat{\tns{\psi}}_{12}}
\newcommand{\drotmultTildei}{\hat{\tns{\theta}}_{i}}
\newcommand{\drotmultTildeA}{\hat{\tns{\theta}}_{1}}
\newcommand{\drotmultTildeB}{\hat{\tns{\theta}}_{2}}

\newcommand{\gtriad}[1]{\tns{g}_{#1}}

\newcommand{\rotvec}{\tns{\psi}}

\newcommand{\rotvecaxis}{\tns{e}_{\psi}}
\newcommand{\rotvecnorm}{\psi}
\newcommand{\drotvec}{\delta\rotvec}
\newcommand{\rotmult}{\tns{\theta}}
\newcommand{\rotmulti}{\tns{\theta}_i}
\newcommand{\rotmultA}{\tns{\theta}_1}
\newcommand{\rotmultB}{\tns{\theta}_2}
\newcommand{\drotmult}{\delta\rotmult}
\newcommand{\drotmultA}{\delta\rotmultA}
\newcommand{\drotmultB}{\delta\rotmultB}
\newcommand{\drotmulti}{\delta\tns{\theta}_i}

\newcommand{\Drotmulti}{\Delta\rotmulti}

\newcommand{\objectiveVariation}{\delta_{o}}

\newcommand{\SskewOp}{\tnss{S}}
\newcommand{\Sskew}[1]{\SskewOp{\br{#1}}}
\newcommand{\Ttrans}{\tnss{T}}
\newcommand{\pen}{\epsilon}

\newcommand{\dWLagrange}{\delta W^\lambda}

\newcommand{\constraintCouplingBase}{\mat{Q}}

\newcommand{\rbeamO}{\tns{r}_0}
\newcommand{\rbeam}{\tns{r}}
\newcommand{\rbeami}{\tns{r}_i}
\newcommand{\rbeamA}{\tns{r}_1}
\newcommand{\rbeamB}{\tns{r}_2}
\newcommand{\rbeamAO}{\tns{r}^0_1}
\newcommand{\rbeamBO}{\tns{r}^0_2}
\newcommand{\drbeami}{\delta\rbeami}
\newcommand{\drbeamA}{\delta\rbeamA}
\newcommand{\drbeamB}{\delta\rbeamB}
\newcommand{\Drbeami}{\Delta\rbeami}

\newcommand{\ubeam}[1][]{\tns{u}_{#1}}

\newcommand{\radiusAvg}{\bar{R}}
\newcommand{\radiusA}{R_1}
\newcommand{\radiusB}{R_2}

\newcommand{\density}{\rho}

\newcommand{\rbeamBA}{\tns{r}_{21}}

\newcommand{\rbeamBAO}{\tns{r}^0_{21}}
\newcommand{\RbeamBAOi}{{}^i\tns{R}^0_{21}}
\newcommand{\RbeamBAOA}{{}^1\tns{R}^0_{21}}
\newcommand{\RbeamBAOB}{{}^2\tns{R}^0_{21}}
\newcommand{\RbeamABOA}{{}^1\tns{R}^0_{12}}
\newcommand{\RbeamABOB}{{}^2\tns{R}^0_{12}}

\newcommand{\RbeamBAi}{{}^i\tns{R}_{21}}

\newcommand{\rotvecA}{\rotvec{}_1}
\newcommand{\rotvecB}{\rotvec{}_2}

\newcommand{\placeholderPos}{r}
\newcommand{\placeholderRot}{\theta}
\newcommand{\dWGen}{\delta W^\lambda_{\placeholder}}
\newcommand{\dWGenStar}{\delta W^{\lambda\ast}_{\placeholder}}
\newcommand{\dWPosStar}{\delta W^{\lambda\ast}_{\placeholderPos}}
\newcommand{\dWRotStar}{\delta W^{\lambda\ast}_{\placeholderRot}}
\newcommand{\dWPenGen}{\delta W^{\epsilon}_{\placeholder}}
\newcommand{\dWPos}{\delta W^\lambda_{\placeholderPos}}
\newcommand{\dWRot}{\delta W^\lambda_{\placeholderRot}}
\newcommand{\PiGen}{\Pi^\lambda_{\placeholder}}
\newcommand{\PiPenGen}{\Pi^\epsilon_{\placeholder}}
\newcommand{\dlagrangeGen}{\delta \tns{\lambda}_{\placeholder}}
\newcommand{\dlagrangePos}{\delta \tns{\lambda}_{\placeholderPos}}
\newcommand{\dlagrangeRot}{\delta \tns{\lambda}_{\placeholderRot}}
\newcommand{\DlagrangeGen}{\Delta \tns{\lambda}_{\placeholder}}
\newcommand{\DlagrangePos}{\Delta \tns{\lambda}_{\placeholderPos}}
\newcommand{\DlagrangeRot}{\Delta \tns{\lambda}_{\placeholderRot}}
\newcommand{\lagrangeGen}{\tns{\lambda}_{\placeholder}}
\newcommand{\lagrangePos}{\tns{\lambda}_{\placeholderPos}}
\newcommand{\lagrangeRot}{\tns{\lambda}_{\placeholderRot}}
\newcommand{\couplingGen}{\equationAndInline{\tns{g}_{\placeholder}}{\tns{g}{}_{\placeholder}}}
\newcommand{\couplingGenO}{\tns{g}^0_{\placeholder}}
\newcommand{\couplingPos}{\tns{g}_{\placeholderPos}}

\newcommand{\couplingRot}{\tns{g}_{\placeholderRot}}

\newcommand{\deformationPosMati}{{}^i\hat{\tns{R}}_{21}}
\newcommand{\deformationPosMatA}{{}^1\hat{\tns{R}}_{21}}
\newcommand{\deformationPosMatB}{{}^2\hat{\tns{R}}_{21}}
\newcommand{\deformationPos}{\hat{\tns{r}}_{21}}

\newcommand{\deformationPosSym}{\hat{\tns{r}}_{12}}

\newcommand{\rigidBodyTranslation}{\tns{a}}
\newcommand{\rigidBodyRotation}{\triad^{\ast}}

\newcommand{\rBA}{\vv{f}}
\newcommand{\rBAi}{\rBA_i}
\newcommand{\rBAA}{\rBA_1}
\newcommand{\rBAB}{\rBA_2}

\newcommand{\rBAPenGeni}{\rBA^\epsilon_{i,\placeholder}}
\newcommand{\rBAPenGenA}{\rBA^\epsilon_{1,\placeholder}}
\newcommand{\rBAPenGenB}{\rBA^\epsilon_{2,\placeholder}}
\newcommand{\rBAgeni}{\rBA^\lambda_{i,\placeholder}}
\newcommand{\rBAgenA}{\rBA^\lambda_{1,\placeholder}}
\newcommand{\rBAgenB}{\rBA^\lambda_{2,\placeholder}}

\newcommand{\dqi}{\delta\vv{q}_i}
\newcommand{\Dqi}{\Delta\vv{q}_i}

\newcommand{\dqA}{\delta\vv{q}_1}
\newcommand{\DqA}{\Delta\vv{q}_1}

\newcommand{\dqB}{\delta\vv{q}_2}
\newcommand{\DqB}{\Delta\vv{q}_2}

\newcommand{\penGen}{\pen_{\placeholder}}
\newcommand{\penPos}{\pen_{\placeholderPos}}
\newcommand{\penRot}{\pen_{\placeholderRot}}

\newcommand{\constraintCouplingQiLGen}{\constraintCouplingBase^{q_i\lambda}_{\placeholder}}
\newcommand{\constraintCouplingQALGen}{\constraintCouplingBase^{q_1\lambda}_{\placeholder}}
\newcommand{\constraintCouplingQBLGen}{\constraintCouplingBase^{q_2\lambda}_{\placeholder}}
\newcommand{\constraintCouplingLQiGen}{\constraintCouplingBase^{\lambda q_i}_{\placeholder}}
\newcommand{\constraintCouplingLQAGen}{\constraintCouplingBase^{\lambda q_1}_{\placeholder}}
\newcommand{\constraintCouplingLQBGen}{\constraintCouplingBase^{\lambda q_2}_{\placeholder}}
\newcommand{\constraintCouplingQiQjGen}{\constraintCouplingBase^{q_iq_j}_{\placeholder}}
\newcommand{\constraintCouplingQAQAGen}{\constraintCouplingBase^{q_1q_1}_{\placeholder}}
\newcommand{\constraintCouplingQBQBGen}{\constraintCouplingBase^{q_2q_2}_{\placeholder}}
\newcommand{\constraintCouplingQAQBGen}{\constraintCouplingBase^{q_1q_2}_{\placeholder}}
\newcommand{\constraintCouplingQBQAGen}{\constraintCouplingBase^{q_2q_1}_{\placeholder}}

\newcommand{\constraintCouplingQALPos}{\constraintCouplingBase^{q_1\lambda}_{\placeholderPos}}
\newcommand{\constraintCouplingQBLPos}{\constraintCouplingBase^{q_2\lambda}_{\placeholderPos}}
\newcommand{\constraintCouplingQAQAPos}{\constraintCouplingBase^{q_1q_1}_{\placeholderPos}}
\newcommand{\constraintCouplingQBQBPos}{\constraintCouplingBase^{q_2q_2}_{\placeholderPos}}
\newcommand{\constraintCouplingQAQBPos}{\constraintCouplingBase^{q_1q_2}_{\placeholderPos}}
\newcommand{\constraintCouplingQBQAPos}{\constraintCouplingBase^{q_2q_1}_{\placeholderPos}}
\newcommand{\constraintCouplingLQAPos}{\constraintCouplingBase^{\lambda q_1}_{\placeholderPos}}
\newcommand{\constraintCouplingLQBPos}{\constraintCouplingBase^{\lambda q_2}_{\placeholderPos}}

\newcommand{\constraintCouplingQALRot}{\constraintCouplingBase^{q_1\lambda}_{\placeholderRot}}
\newcommand{\constraintCouplingQBLRot}{\constraintCouplingBase^{q_2\lambda}_{\placeholderRot}}
\newcommand{\constraintCouplingQAQARot}{\constraintCouplingBase^{q_1q_1}_{\placeholderRot}}
\newcommand{\constraintCouplingQBQBRot}{\constraintCouplingBase^{q_2q_2}_{\placeholderRot}}
\newcommand{\constraintCouplingQAQBRot}{\constraintCouplingBase^{q_1q_2}_{\placeholderRot}}
\newcommand{\constraintCouplingQBQARot}{\constraintCouplingBase^{q_2q_1}_{\placeholderRot}}
\newcommand{\constraintCouplingLQARot}{\constraintCouplingBase^{\lambda q_1}_{\placeholderRot}}
\newcommand{\constraintCouplingLQBRot}{\constraintCouplingBase^{\lambda q_2}_{\placeholderRot}}

\newcommand{\qbeami}{\vv{d}_i}
\newcommand{\qbeamA}{\vv{d}_1}
\newcommand{\qbeamB}{\vv{d}_2}

\newcommand{\dqbeami}{\delta\qbeami}
\newcommand{\dqbeamA}{\delta\qbeamA}
\newcommand{\dqbeamB}{\delta\qbeamB}

\newcommand{\Dqbeami}{\Delta\qbeami}
\newcommand{\DqbeamA}{\Delta\qbeamA}
\newcommand{\DqbeamB}{\Delta\qbeamB}
\newcommand{\Hbeami}{\mat{H}_i}
\newcommand{\HbeamA}{\mat{H}_1}
\newcommand{\HbeamB}{\mat{H}_2}

\newcommand{\DHbeamLini}{\mat{H}^{\Delta}_i}
\newcommand{\DHbeamLinA}{\mat{H}^{\Delta}_1}
\newcommand{\DHbeamLinB}{\mat{H}^{\Delta}_2}

\newcommand{\Nbeami}{\mat{N}_i}
\newcommand{\NbeamA}{\mat{N}_1}
\newcommand{\NbeamB}{\mat{N}_2}
\newcommand{\xiBeami}{\xi_i}

\newcommand{\nDofBeami}{n_{i}}
\newcommand{\nDofBeamA}{n_{A}}
\newcommand{\nDofBeamB}{n_{B}}

\theoremstyle{thmstyleone}%
\theoremstyle{thmstyletwo}%

\theoremstyle{thmstylethree}%

\begin{document}

\title{A variationally consistent beam-to-beam point coupling formulation for geometrically exact beam theories}


\author*[1]{\fnm{Ivo} \sur{Steinbrecher}}\email{ivo.steinbrecher@unibw.de}

\author[1,2]{\fnm{Nora} \sur{Hagmeyer}}

\author[3]{\fnm{Christoph} \sur{Meier}}

\author[1]{\fnm{Alexander} \sur{Popp}}

\affil[1]{\orgdiv{Institute for Mathematics and Computer-Based Simulation (IMCS)}, \orgname{University of the Bundeswehr Munich}, \orgaddress{\street{Werner-Heisenberg-Weg 39, D-85577}, \city{Neubiberg}, \postcode{85577}, \country{Germany}}}

\affil[2]{\orgname{Daisytuner}, \orgaddress{\street{Am Mehlstück 14}, \city{Wetzlar}, \postcode{35583}, \country{Germany}}}

\affil[3]{\orgdiv{Professorship of Simulation for Additive Manufacturing Processes (SAM)}, \orgname{Technical University of Munich}, \orgaddress{\street{Freisinger Landstraße 52}, \city{Garching b. München}, \postcode{85748}, \country{Germany}}}


\abstract{%
Slender beam-like structures frequently occur in engineering applications and often interact at discrete locations through joints or connectors. 
Accurate modeling of such interactions is particularly challenging when different numerical formulations are involved in terms of underlying beam theory, interpolation schemes, and rotation parametrization.
In this work, a versatile formulation-independent beam-to-beam point coupling approach is proposed within the framework of the geometrically exact beam theory discretized by the finite element method.
The coupling constraints are expressed solely in terms of cross-section kinematics, namely centroid positions and orientations.
Suitable generalized deformation measures for positional and rotational coupling are introduced, allowing for general coupling configurations, including relative rotations and non-coincident cross-section centroids in the reference configuration.
The contribution of the coupling conditions to the weak form of the balance equations is derived in a variationally consistent manner and can be incorporated directly into the weak form of existing beam finite element models.
Constraint enforcement is formulated using a Lagrange multiplier method and a penalty regularization.
The proposed approach satisfies key properties such as objectivity, symmetry, and consistency with a stress-free reference configuration.
Numerical examples demonstrate the robustness and flexibility of the method for coupling beams with different formulations and discretizations, even when the interaction points are located at arbitrary positions within beam elements.
}




\maketitle

\section{Introduction}

Slender fiber- or rod-like components are key elements in many engineering applications and scientific fields, including fiber-reinforced structures, lattice and truss systems, flexible mechanisms, and biological filaments.
Such slender structures can be efficiently modeled using one-dimensional Cosserat continua, based on structural beam theories.
In many practical applications, these components interact with each other at discrete locations, forming joints or connectors that play an essential role in the overall mechanical behavior of the system.
Accurate and robust modeling of such couplings is therefore essential, particularly when interacting beams are described by different beam formulations or discretization schemes.

In this work, we consider beam models based on the geometrically exact beam theory.
The theoretical groundwork for the geometrically exact \sr beam theory was laid in~\cite{Reissner1972,Reissner1981}, the first finite element implementation was presented in~\cite{Simo1988} and has subsequently been further developed and improved.
In~\cite{Jelenic1999,Crisfield1999}, objective strain measures and corresponding objective finite element discretizations were introduced for geometrically exact beam formulations.
A director based rotational interpolation scheme for geometrically exact beams was presented in~\cite{Betsch2003}.
An efficient~$SE(3)$ interpolation for a uniform treatment of beam centerline and orientation fields was proposed in~\cite{Sonneville2014}.
All of these works consider shear-deformable beam formulations.
A shear-rigid (\kl) geometrically exact beam formulation was developed in~\cite{Meier2014,Meier2015}.
More recently, a mixed geometrically exact beam formulation was proposed in~\cite{Herrmann2026}, which allows for a unified formulation that can model shear-deformable and shear-rigid beams as well as extensible and inextensible beams.

Despite their differences, these beam formulations share a common kinematic description: each beam cross-section is uniquely defined by its centroid position~$\rbeam \in \R{3}$ and its orientation~$\triad \in \SO$.
Here,~$\triad = [\gtriad{1},\gtriad{2},\gtriad{3}]$ denotes an orthonormal triad formed by the cross-section base vectors~$\gtriad{i} \in \R{3}$.
While the underlying kinematic variables are identical, their parametrization and finite element interpolation differ between beam formulations.

Rigid joints between beams have been considered in several of the previously mentioned works by directly coupling the nodal degrees of freedom of the interacting beams~\cite{Simo1988,Jelenic1999,Sonneville2014}.
However, this approach is limited to coupling beams at finite element nodes.
Furthermore, when more complex parametrizations of the orientation field are used, as in \kl beam formulations, such direct coupling becomes less straightforward.
An alternative approach is to formulate explicit coupling constraints in terms of the cross-section kinematics.
Typically, three positional constraint equations enforce the coincidence of the centroid positions, while three rotational constraint equations enforce the relative orientation of the cross-section frames.
For the latter, a commonly used approach is to enforce the orthogonality of selected pairs of base vectors of the two cross-sections.
Although this approach is valid, it may introduce numerical difficulties.
For instance, if the orientational constraints are enforced using a penalty method, the resulting system matrix may lose positive definiteness, which can lead to severe convergence issues in nonlinear solution schemes~\cite[Section~3.7.2]{Vetyukov2014}.

In this work, we follow a different strategy to formulate the coupling constraints, which is inspired by the framework proposed in~\cite{Meier2023}.
In that work, general interaction potentials between beams were introduced within the geometrically exact beam theory.
This framework has already been successfully applied to model potential-based beam interactions~\cite{Grill2020a,Grill2021} and the coupling of beams with three-dimensional continua~\cite{Steinbrecher2020,Steinbrecher2022,Steinbrecher2025}.
In~\cite{Meier2023}, the authors also outlined how two beam cross-sections can be coupled to each other.
In the present work, we build upon this approach and modify the coupling constraints to ensure symmetry with respect to the interacting cross-sections.
The main idea is to introduce suitable \emph{generalized deformation measures} for positional and rotational coupling that describe the relative deformation between interacting beam cross-sections.
The formulation is designed to handle general coupling configurations in which the cross-sections may exhibit relative rotations in the reference configuration and their centroid positions do not necessarily coincide.
Also, the coupling point is not restricted to be located at a finite element node, but can be located at an arbitrary position along the beam elements.
We show that the proposed coupling formulation satisfies fundamental properties such as objectivity, symmetry, and consistency with a stress-free reference configuration.
Furthermore, we present the discrete coupling equations in a general form such that they can be applied to arbitrary geometrically exact beam formulations, independent of the choice of primary variables, rotational parametrizations, or interpolation schemes.
As a result, the proposed formulation provides a robust and formulation-independent framework for coupling geometrically exact beam models in complex beam assemblies.

The remainder of this paper is organized as follows.
In \Cref{sec:large_rotations}, we briefly introduce the relevant concepts of large rotations needed for the formulation of the coupling constraints.
\Cref{sec:governing_equations} introduces the variational formulation of the beam-to-beam coupling constraints.
The discrete implementation of the coupling equations is presented in \Cref{sec:discretized_coupling_equations}.
Numerical examples demonstrating the accuracy and robustness of the method are presented in \Cref{sec:examples}.
Finally, Appendix~\ref{sec:appendix_properties_positional_coupling} demonstrates objectivity and the other fundamental properties of the coupling constraints.

\section{Large rotations}
\label{sec:large_rotations}

This section briefly summarizes the notation and fundamental relations for finite rotations required in this work.
For a more comprehensive treatment, the reader is referred to the literature, \eg,~\cite{Cardona1988,Ibrahimbegovic1995,Romero2004,Crisfield1997,Meier2019, Betsch1998}.

A rotation can be described by an orthonormal triad
\begin{equation}
\triad = \matrix{\gtriad{1}, \gtriad{2}, \gtriad{3}} \in \SO,
\end{equation}
where $\SO$ denotes the special orthogonal group, i.e., the triad satisfies~$\triad\tr \triad = \tnssI$ and~$\det\br{\triad} = 1$.
The columns~$\gtriad{i}$ represent the basis vectors of the rotated configuration.

For parametrization, a rotation vector~$\rotvec \in \R{3}$ is introduced such that $\triad = \triad(\rotvec)$.
The vector~$\rotvec$ defines a rotation with magnitude $\rotvecnorm = \norm{\rotvec}$ about the axis $\rotvecaxis = \rotvec / \rotvecnorm$.
The corresponding rotation tensor is obtained via the exponential map
\begin{equation}
\triad(\rotvec) = \exp\!\big(\Sskew{\rotvec}\big),
\end{equation}
which can be expressed using Rodrigues' formula as
\begin{equation}
\triad(\rotvec)
= \tnssI
+ \frac{\sin \rotvecnorm}{\rotvecnorm} \Sskew{\rotvec}
+ \frac{1 - \cos \rotvecnorm}{\rotvecnorm^2} \Sskew{\rotvec}^2.
\end{equation}
Here, $\Sskew{\cdot}$ denotes the skew-symmetric operator associated with the cross product, i.e.,~$\Sskew{\tns{a}}\tns{b} = \tns{a} \times \tns{b} \, \forall \, \tns{a}, \tns{b} \in \R{3}$.
The inverse mapping is denoted by
\begin{equation}
\label{eq:logSO3}
\rotvec = \rv(\triad),
\end{equation}
which extracts the rotation vector from a given rotation tensor.

Given two rotations $\triad_1$ and $\triad_2$, the relative rotation between them is defined by
\begin{equation}
\triad_{21} = \triad_2 \triad_1\tr,
\end{equation}
with the corresponding rotation vector $\rotvec_{21} = \rv(\triad_{21})$.
Due to the nonlinear structure of the rotation group, this relative rotation vector does not satisfy $\rotvec_{21} = \rotvec_2 - \rotvec_1$.

Variations of the rotation tensor can be expressed using a multiplicative increment in terms of the spin vector~$\drotmult \in \R{3}$,
\begin{equation}
\delta \triad = \Sskew{\drotmult}\,\triad.
\end{equation}
% Consequently, the variations of the triad basis vectors are given by
% \begin{equation}
% \dgtriad{i} = \drotmult \times \gtriad{i}.
% \end{equation}
The variation of the rotation tensor can also be expressed in terms of additive variations of the rotation vector~$\drotvec \in \R{3}$,
\begin{equation}
\delta \triad = \pfrac{\triad}{\rotvec} \drotvec
.
\end{equation}
Here, the transformation matrix~$\Ttrans \in \R{3\times 3}$ maps the multiplicative variations to the additive variations, i.e.,
\begin{equation}
\drotvec = \Ttrans(\rotvec)\, \drotmult
.
\end{equation}
A closed-form expression reads
\begin{equation}
\Ttrans(\rotvec) =
\frac{1}{\rotvecnorm^2} \rotvec \rotvec\tr
- \frac{1}{2} \Sskew{\rotvec}
+ \frac{\rotvecnorm}{2 \tan(\rotvecnorm/2)}
  \left( \tnssI - \frac{1}{\rotvecnorm^2} \rotvec \rotvec\tr \right).
\end{equation}

\section{Coupling equations}
\label{sec:governing_equations}

In this section, we present the governing equations of the proposed beam-to-beam point coupling formulation.
We consider two cross-sections~$1$ and~$2$ that are coupled to each other.
It is sufficient to consider a single cross-section pair, since the formulation generalizes straightforwardly to an arbitrary number of coupling pairs.
The kinematics of both cross-sections are fully described by their centroid positions~$\rbeamA, \rbeamB \in \R{3}$ and orientations~$\triadA, \triadB \in \SO$ in the current configuration.
The corresponding quantities in the reference configuration are denoted by~$\rbeamAO, \rbeamBO \in \R{3}$ and~$\triadAO, \triadBO \in \SO$.
In this work, a superscript~$0$ denotes quantities in the reference configuration.
In the following, we derive the variational formulation of the coupling constraints in terms of these kinematic quantities and their variations.
The resulting virtual work contributions can be added directly to the weak formulation of the underlying beam model.
Consequently, the presented coupling formulation is independent of the specific beam theory employed and can be combined with any geometrically exact beam formulation, such as the \sr or \kl beam theory.

%In the following, we first introduce the general formulation of the coupling constraints and present the enforcement of the constraints using a Lagrange multiplier formulation and a penalty regularization.
%Then, the positional and rotational coupling constraints are introduced.

\subsection{Enforcement of coupling constraints}

Before specifying the explicit forms of the positional and rotational constraints, we briefly introduce the enforcement of coupling constraints employed in this work in a generic manner.
This applies to both positional and rotational types and will therefore not be repeated in the subsequent subsections.
We state the constraint equations in the following general form:
\begin{align}
\label{eq:general_coupling_constraint}
\couplingGen\br{\rbeamA, \triadA, \rbeamB, \triadB} = \tnsO
,
\end{align}
where~$\placeholder$ is a placeholder for either the positional or rotational coupling constraint, which will be denoted by~$\placeholderPos$ and~$\placeholderRot$, respectively, in the following sections, cf. \Cref{sec:positional_coupling,sec:rotational_coupling}.
Each constraint is vector-valued with~$\couplingGen \in \R{3}$.
Consequently, a total of six scalar constraint equations restrict the six relative degrees of freedom between the two cross-sections.

\subsubsection{Lagrange multiplier method}

In this subsection, we describe the enforcement of the coupling constraints using the Lagrange multiplier method, which provides a consistent variational framework for deriving the generalized coupling equations.
For each vector-valued constraint, a Lagrange multiplier vector~$\lagrangeGen \in \R{3}$ is introduced.
The associated Lagrange multiplier potential is defined as
\begin{align}
\PiGen
= \lagrangeGen\tr \couplingGen
,
\end{align}
where the explicit dependence of the coupling constraints on the kinematic variables is omitted for readability.
Taking the variation of the Lagrange multiplier potential yields the virtual work contribution of the coupling constraints,
\begin{align}
\label{eq:lagrange_multiplier_weak_form}
\dWGen
=
\delta \PiGen
=
\dlagrangeGen\tr \couplingGen
+
\lagrangeGen\tr \objectiveVariation \couplingGen
.
\end{align}
Here, the variation of the coupling constraints is expressed in terms of objective variations, which is required to ensure the objectivity of the formulation, see condition~\eqref{item:objectivity} below.
The virtual work contribution~\eqref{eq:lagrange_multiplier_weak_form} can be written in matrix-vector form as
\begin{align}
\label{eq:lagrange_multiplier_weak_form_matrix_vector}
\dWGen =
\dlagrangeGen\tr \couplingGen
+
\dqA\tr \underbrace{\constraintCouplingQALGen \lagrangeGen}_{\rBAgenA}
+
\dqB\tr \underbrace{\constraintCouplingQBLGen \lagrangeGen}_{\rBAgenB}
.
\end{align}
Here,~$\constraintCouplingQiLGen \in \R{6\times 3}$ are coupling matrices that map the Lagrange multiplier vector to the corresponding generalized force and moment contributions~$\rBAgeni \in \R{6}$ acting on the cross-sections, i.e.,~$\rBAgeni = \constraintCouplingQiLGen \lagrangeGen$.
Moreover, the vector~$\dqi \in \R{6}$ collects the variations of all kinematic variables of a single cross-section, i.e.,
\begin{align}
\dqi =
\vector{
    \drbeami\tr &
    \drotmulti\tr
}\tr
.
\end{align}
Finally, we provide the linearization of the generalized forces, which is required for the consistent incorporation of the coupling constraints into tangent-based nonlinear solution schemes.
The linearization reads
\begin{align}
\label{eq:saddle_point_system}
\Delta
\vector{
    \rBAgenA \\
    \rBAgenB \\
    \couplingGen
}
=
\matrix{
    \constraintCouplingQAQAGen & \constraintCouplingQAQBGen & \constraintCouplingQALGen \\
    \constraintCouplingQBQAGen & \constraintCouplingQBQBGen & \constraintCouplingQBLGen \\
    \constraintCouplingLQAGen & \constraintCouplingLQBGen & \mat{0}
}
\vector{
    \DqA \\
    \DqB \\
    \DlagrangeGen
}
.
\end{align}
Here,~$\constraintCouplingQiQjGen \in \R{6\times 6}$ denote the linearization of the generalized forces with respect to the kinematic variables, while~$\constraintCouplingLQiGen \in \R{3\times 6}$ represents the linearization of the coupling constraints with respect to the kinematic variables.
Furthermore,~$\DlagrangeGen \in \R{3}$ denotes the increment of the Lagrange multipliers and~$\Dqi \in \R{6}$ collects the increments of the kinematic variables, i.e.,
\begin{align}
\Dqi =
\vector{
    \Drbeami\tr &
    \Drotmulti\tr
}\tr
.
\end{align}

\subsubsection{Penalty regularization}

The linear system resulting from the Lagrange multiplier formulation~\eqref{eq:saddle_point_system} contains the Lagrange multipliers as additional unknowns and exhibits saddle-point structure.
To avoid these additional complexities, a penalty regularization can be employed to eliminate the Lagrange multipliers from the system.
We employ a simple penalty regularization of the form
\begin{align}
\label{eq:penalty_regularization}
    \lagrangeGen \approx \penGen \couplingGen\br{\rbeamA, \triadA, \rbeamB, \triadB}
,
\end{align}
where~$\penGen \in \R{+}$ denotes a scalar penalty parameter.
For~$\penGen \rightarrow \infty$, the penalty regularization~\eqref{eq:penalty_regularization} becomes equivalent to the original constraint equations~\eqref{eq:general_coupling_constraint}.
With the penalty regularization, the Lagrange multipliers are no longer explicit unknowns and the variational form~\eqref{eq:lagrange_multiplier_weak_form_matrix_vector} can be written as
\begin{align}
\label{eq:penalty_regularized_weak_form}
\dWPenGen =
\dqA\tr
\underbrace{
\penGen
\constraintCouplingQALGen \couplingGen
}_{\rBAPenGenA}
+
\dqB\tr
\underbrace{
\penGen
\constraintCouplingQBLGen \couplingGen
}_{\rBAPenGenB}
.
\end{align}
Here,~$\rBAPenGeni$ denote the generalized penalty forces and moments resulting from the coupling constraints.
The linearization of the generalized penalty contributions reads
\begin{align}
\Delta
\vector{
    \rBAPenGenA
    \\
    \rBAPenGenB
}
=
\br{
\matrix{
    \constraintCouplingQAQAGen & \constraintCouplingQAQBGen \\
    \constraintCouplingQBQAGen & \constraintCouplingQBQBGen \\
}
+
\penGen
\matrix{
    \constraintCouplingQALGen \constraintCouplingLQAGen & \constraintCouplingQALGen \constraintCouplingLQBGen \\
    \constraintCouplingQBLGen \constraintCouplingLQAGen & \constraintCouplingQBLGen \constraintCouplingLQBGen \\
}
}
\vector{
    \DqA \\
    \DqB
}
.
\end{align}

The penalty parameters~$\penGen$ control the enforcement of the constraints when a penalty regularization is employed.
In practice, they should be chosen sufficiently large to enforce the constraints accurately, while avoiding ill-conditioning of the resulting system of equations.
Suitable values for the penalty parameters can be estimated based on the stiffness and geometry of the beams.
As a rule of thumb, the positional penalty parameter may be chosen as~$\penPos \approx \EbeamAvg \radiusAvg$,
where~$\EbeamAvg=\tfrac{1}{2}(\EbeamA+\EbeamB)$ and~$\radiusAvg=\tfrac{1}{2}(\radiusA+\radiusB)$ denote the average Young's modulus and cross-section size of the two beams, respectively.
For the rotational penalty parameter, a suitable choice is~$\penRot \approx \EbeamAvg \radiusAvg^3$.
These estimates depend on the specific problem and may require some tuning, but they provide a good starting point for the selection of the penalty parameters.

\begin{remark}
    The presented penalty regularization can alternatively be derived from a quadratic penalty potential of the form~$\PiPenGen = \frac{1}{2} \penGen (\couplingGen)\tr \couplingGen$.
\end{remark}

\begin{remark}
    The presented derivation of the penalty regularized system of equations can also be interpreted in an algebraic way, cf.~\cite{Zienkiewicz2013}, by adding the term~$-\dlagrangeGen\tr \frac{1}{\penGen} \lagrangeGen$ to the variational formulation~\eqref{eq:lagrange_multiplier_weak_form}, which vanishes for~$\penGen \rightarrow \infty$.
    By adding this term, the Lagrange multipliers can be condensed from the last row of~\eqref{eq:lagrange_multiplier_weak_form_matrix_vector}, yielding~\eqref{eq:penalty_regularized_weak_form}.
\end{remark}

\begin{remark}
    In this work, a simple penalty regularization with a scalar penalty parameter is considered.
    The general framework for employing a symmetric material penalty tensor has been outlined in~\cite{Meier2023}.
    Incorporating such a tensor into the present formulation would allow the modeling of complex geometric and material behavior of the connection between the two cross-sections.
    However, this extension is beyond the scope of the present work.
\end{remark}

\subsection{Requirements for coupling constraints}
\label{sec:requirements_coupling_constraints}

In order to ensure mechanical consistency and frame-independence of the formulation, the coupling constraints are required to satisfy the following fundamental properties:
\begin{enumerate}[(i)]

\item \label{item:stress_free}\emph{Stress-free reference configuration:}
The coupling constraints must vanish in the reference configuration, i.e.,
\begin{align}
\label{eq:reference_requirement}
\couplingGenO
=
\couplingGen\br{\rbeamAO, \triadAO, \rbeamBO, \triadBO}
=
\tnsO
.
\end{align}

\item \label{item:symmetry}\emph{Symmetry:}
A permutation of the cross-section indices~$1$ and~$2$ shall only lead to a sign change of the constraint equations,
\begin{align}
\label{eq:symmetry_requirement}
\couplingGen\br{\rbeamA, \triadA, \rbeamB, \triadB}
=
- \couplingGen\br{\rbeamB, \triadB, \rbeamA, \triadA}
.
\end{align}

\item \label{item:objectivity}\emph{Objectivity:}
The variational formulation must be invariant under superimposed rigid body motions.
For an arbitrary rigid body motion defined by a constant rotation~$\rigidBodyRotation \in \SO$ and translation~$\rigidBodyTranslation \in \R{3}$, the kinematic variables, Lagrange multipliers, and variations transform as
\begin{equation}
\begin{aligned}
\label{eq:rigid_body_transformations}
\rbeami^\ast = \rigidBodyRotation \rbeami + \rigidBodyTranslation
,
\quad
\triadi^\ast = \rigidBodyRotation \triadi
,
\quad
\lagrangeGen^\ast = \rigidBodyRotation \lagrangeGen
,
\quad
\drbeami^\ast = \rigidBodyRotation \drbeami
,
\quad
\drotmulti^\ast = \rigidBodyRotation \drotmulti
\quad
\text{and}
\quad
\dlagrangeGen^\ast = \rigidBodyRotation \dlagrangeGen
.
\end{aligned}
\end{equation}
To guarantee objectivity, the virtual work contribution of the Lagrange multiplier potential~\eqref{eq:lagrange_multiplier_weak_form} must satisfy
\begin{equation}
\begin{aligned}
\dWGen
&=
\dWGen\br[1]{
    \rbeamA, \triadA, \rbeamB, \triadB, \drbeamA, \drotmultA, \drbeamB, \drotmultB, \dlagrangeGen
}
\\
&= \dWGen\br[1]{
    \rbeamA^\ast, \triadA^\ast, \rbeamB^\ast, \triadB^\ast, \drbeamA^\ast, \drotmultA^\ast, \drbeamB^\ast, \drotmultB^\ast, \dlagrangeGen^\ast
}
=
\dWGenStar
.
\end{aligned}
\end{equation}

\end{enumerate}
The fulfillment of these requirements will be demonstrated for the positional and rotational constraints in the subsequent sections.

\begin{remark}
The symmetry requirement~\eqref{item:symmetry} stated above is only relevant for the penalty regularization.
In the Lagrange multiplier case, the constraint equations are satisfied exactly, i.e.,~$\couplingGen = \tnsO$, such that a permutation of the cross-section indices yields the same result.
In contrast, when the constraint equations are enforced with a penalty regularization, a small deviation from the solution with exact constraint enforcement will occur, i.e.,~$\couplingGen \neq \tnsO$.
In this case, the symmetry requirement becomes essential to ensure that the formulation does not depend on the ordering of the two cross-sections.
\end{remark}

\subsection{Positional coupling}
\label{sec:positional_coupling}

In this section, we present the positional coupling between the two cross-sections~$1$ and~$2$.
The positional constraint enforces the coincidence of selected material points of the two cross-sections.
Since these material points are not necessarily located at the centroids of the cross-sections, the positional constraint generally introduces coupling moments.
As a first step, we define the relative distance vector~$\rbeamBA$ between the centroids of the two cross-sections:
\begin{align}
\rbeamBA = \rbeamB - \rbeamA
.
\end{align}
The relative distance vector~$\rbeamBA\in\R{3}$ constitutes a \emph{generalized spatial deformation measure}~\cite{Meier2023}.
By introducing suitable push-forward and pull-back operators, a material representation of~$\rbeamBA$ can be obtained.
The definition of these operators corresponds to the choice of a reference frame in which the relative distance vector is expressed.
At this stage, the appropriate reference frame for the material representation is not yet specified.
Therefore, we formulate the push-forward and pull-back operations with respect to both cross-section frames and denote them by~$\triad_i$ and~$\triad_i\tr$, respectively, with~$i=1,2$.
The specific choice will become apparent once the explicit form of the positional deformation measures is introduced.
The material relative distance vector can be stated as
\begin{align}
\RbeamBAi = \triad_i\tr \rbeamBA
.
\end{align}
In this work, a leading superscript is employed to identify the reference frame in which a given quantity is expressed.
The reference material distance vector is defined accordingly as~$\RbeamBAOi = \br[0]{\triad^0_i}\tr \rbeamBAO$.
The material deformation measure associated with the positional coupling constraint is defined as the difference between the current and reference material distance vectors,
\begin{align}
\label{eq:material_deformation_measure_pos}
\deformationPosMati = \RbeamBAi - \RbeamBAOi
.
\end{align}
To obtain a spatial deformation measure that is symmetric~\eqref{item:symmetry} with respect to the ordering of the two cross-sections, we take the push-forward of the two material deformation measures~$\deformationPosMatA$ and~$\deformationPosMatB$ and introduce their arithmetic mean.
The resulting spatial deformation measure can thus be stated as
\begin{align}
\label{eq:spatial_deformation_measure_pos}
\deformationPos = \frac{1}{2} \br{\triad_1 \deformationPosMatA + \triad_2 \deformationPosMatB}
=
\rbeamB - \rbeamA -  \frac{1}{2}\br{ \triadA \RbeamBAOA + \triadB \RbeamBAOB }
.
\end{align}
The verification that the deformation measure leads to a coupling formulation fulfilling the fundamental properties~\eqref{item:stress_free}--\eqref{item:objectivity} stated in \Cref{sec:requirements_coupling_constraints} is provided in Appendix~\ref{sec:appendix_properties_positional_coupling}.
Since these requirements are satisfied, the deformation measure can be directly employed as the positional coupling constraint, i.e.,
\begin{align}
\couplingPos = \deformationPos
.
\end{align}
For the variational form arising from the Lagrange multiplier method, cf.~\eqref{eq:lagrange_multiplier_weak_form}, we need to define the objective variation of the spatial deformation measure.
The objective variation is given by the push-forward of the respective material variations~\cite{Meier2023}, i.e.,
\begin{align}
\label{eq:objective_variation_pos}
\objectiveVariation \deformationPos &= \frac{1}{2} \br{\triad_1 \delta\deformationPosMatA + \triad_2 \delta\deformationPosMatB}
.
\end{align}
Here,~$\delta\deformationPosMatA$ and~$\delta\deformationPosMatB$ are the variations of the material deformation measures, which are obtained as
\begin{equation}
\begin{aligned}
\label{eq:variation_material_deformation_measure_pos}
\delta \deformationPosMati = \delta \br{\triad_i\tr \br{\rbeamB - \rbeamA} - \br{\triad^0_i}\tr \br{\rbeamBO - \rbeamAO}} = -\triad_i\tr \Sskew{\drotmulti} \br{\rbeamB - \rbeamA} + \triad_i\tr \br{\drbeamB - \drbeamA}
.
\end{aligned}
\end{equation}
Inserting~\eqref{eq:objective_variation_pos} and~\eqref{eq:variation_material_deformation_measure_pos} into~\eqref{eq:lagrange_multiplier_weak_form} yields the variational form of the positional coupling constraint:
\begin{equation}
\begin{aligned}
\label{eq:weak_form_pos}
\dWPos = \dlagrangePos\tr \br{
    \rbeamB - \rbeamA -  \frac{1}{2}\br{ \triadA \RbeamBAOA + \triadB \RbeamBAOB }
}
+
\lagrangePos\tr \br{
    \drbeamB - \drbeamA
    - \frac{1}{2}\br[1]{ \Sskew{\drotmultB} + \Sskew{\drotmultA}  }  \br{\rbeamB - \rbeamA}
}
.
\end{aligned}
\end{equation}
It is well-known from geometrically exact beam theory that the variation of the cross-section position is work-conjugate to the cross-section force resultants.
Thus, the positional Lagrange multipliers can be interpreted as coupling forces.
Furthermore, the coupling forces induce a moment contribution that is work-conjugate to the cross-section spin vectors.
This moment arises when the cross-section centroids do not coincide in the reference configuration, i.e.,~$\rbeamAO \neq \rbeamBO$.

In a final step, we can express the variational form~\eqref{eq:weak_form_pos} in matrix-vector form~\eqref{eq:lagrange_multiplier_weak_form_matrix_vector} with
\begin{align}
\constraintCouplingQALPos = \constraintCouplingQBLPos =
\vector{
    \tnssI \\
    -\frac{1}{2}\Sskew{\rbeamB - \rbeamA}
}
%    \rbeamB - \rbeamA - \frac{1}{2}\br{ \triadA \RbeamBAOA + \triadB \RbeamBAOB } \\
%    \tns{0}
.
\end{align}
The linearizations read
\begin{align}
\begin{aligned}
\constraintCouplingQAQAPos = \constraintCouplingQBQAPos &= \matrix{
    \tns{0} & \tns{0} \\
    -\frac{1}{2}\Sskew{\lagrangePos} & \tns{0}
}
,
&
\constraintCouplingQAQBPos = \constraintCouplingQBQBPos &= \matrix{
    \tns{0} & \tns{0} \\
    \frac{1}{2}\Sskew{\lagrangePos} & \tns{0}
}
,
\\
\constraintCouplingLQAPos &= \matrix{
    -\tns{I} & \frac{1}{2}\Sskew{\triadA \RbeamBAOA}
},
&
\constraintCouplingLQBPos &= \matrix{
    \tns{I} & \frac{1}{2}\Sskew{\triadB \RbeamBAOB}
}
.
\end{aligned}
\end{align}

\subsection{Rotational coupling}
\label{sec:rotational_coupling}

In this section, we derive the rotational coupling between the two cross-sections.
The rotational coupling enforces a prescribed relative orientation between the cross-section triads.
The formulation of the rotational coupling constraint is based on relative rotations between the two cross-sections, which is in close analogy to the generalized cross-section interaction laws proposed in~\cite{Meier2023}.
We first define the effective rotation that maps the reference cross-section triad to the current configuration,
\begin{align}
\label{eq:effective_rotation}
\triadTildei = \triadi \br{\triadiO}\tr
.
\end{align}
The relative rotation between the effective rotations of both cross-sections is then given by
\begin{align}
\label{eq:relative_rotation_effective}
\triadTildeBA
=
\triadTildeB \triadTildeA\tr
=
\triadB \br{\triadBO}\tr \triadAO \br{\triadA}\tr
.
\end{align}
In~\cite{Meier2023}, it has been shown that the relative rotation (pseudo-)vector associated with~$\triadBA = \triadB \br[0]{\triadA}\tr$ constitutes a suitable \emph{generalized spatial deformation measure} for rotational coupling.
In the present work, we employ a slightly modified definition of the relative rotation vector, which is based on the effective rotations~$\triadTildei$ and therefore directly accounts for the stress-free reference configuration.
The rotation vector can be extracted from~$\triadTildeBA$ with relation~\eqref{eq:logSO3}, i.e.,
\begin{align}
\label{eq:relative_rotation_vector}
\rotvecTildeBA = \rv \br[0]{\triadTildeBA}
.
\end{align}
The verification that the relative rotation vector leads to a coupling formulation fulfilling the fundamental properties~\eqref{item:stress_free}--\eqref{item:objectivity} stated in \Cref{sec:requirements_coupling_constraints} is provided in Appendix~\ref{sec:appendix_properties_rotational_coupling}.
The relative rotation vector~$\rotvecTildeBA$ can therefore be used directly as the rotational coupling constraint, i.e.,
\begin{align}
\couplingRot = \rotvecTildeBA
.
\end{align}
The objective variation of the relative rotation vector~$\objectiveVariation \rotvecTildeBA$ is derived in~\cite{Meier2023} and reads
\begin{align}
\objectiveVariation \rotvecTildeBA = \Ttrans\br[0]{\rotvecTildeBA}\br[1]{\drotmultTildeB - \drotmultTildeA}
.
\end{align}
It remains to relate the spin vectors~$\delta \drotmultTildei$ to the cross-section spin vectors~$\drotmulti$.
Since the reference triads~$\triadiO$ are constant, variation of~\eqref{eq:effective_rotation} yields
\begin{align}
\SskewOp\br[0]{\delta \drotmultTildei} \triadTildei
=
\SskewOp\br{\drotmulti} \triadi \br{\triadiO}\tr
=
\SskewOp\br{\drotmulti} \triadTildei
,
\end{align}
which directly implies
\begin{align}
\delta \drotmultTildei = \drotmulti
.
\end{align}
With these relations, the variational form of the rotational coupling constraint becomes
\begin{align}
\label{eq:weak_form_rot}
\dWRot
=
\dlagrangeRot\tr \rotvecTildeBA
+
\lagrangeRot\tr
\Ttrans\br[0]{\rotvecTildeBA}
\br{\drotmultB - \drotmultA}
.
\end{align}
As multiplicative rotation variations~$\drotmulti$ are work-conjugate to the cross-section moment stress resultants, the associated term~$\pm\Ttrans\tr\br[0]{\rotvecTildeBA}\lagrangeRot$ represents a discrete coupling moment acting between the two beam cross-sections.

In a final step, we can express the variational form~\eqref{eq:weak_form_rot} in matrix-vector form~\eqref{eq:lagrange_multiplier_weak_form_matrix_vector} with
\begin{align}
\constraintCouplingQALRot = \vector{
    \tns{0} \\
    -\Ttrans\tr\br{\rotvecTildeBA}
}
\quad\text{and}\quad
\constraintCouplingQBLRot = \vector{
    \tns{0} \\
    -\Ttrans\tr\br{\rotvecTildeBA}
}
.
\end{align}
The consistent linearization follows as
\begin{align}
\begin{aligned}
\label{eq:jacobian_rotational_coupling}
\constraintCouplingQAQARot = \matrix{
    \tns{0} & \tns{0} \\
    \tns{0} & -\pfrac{\br{\Ttrans\tr\br{\rotvecTildeBA} \lagrangeRot}}{\rotvecA} \Ttrans\br{\rotvecA}
},
\quad
\constraintCouplingQBQBRot = \matrix{
    \tns{0} & \tns{0} \\
    \tns{0} & \pfrac{\br{\Ttrans\tr\br{\rotvecTildeBA} \lagrangeRot}}{\rotvecB} \Ttrans\br{\rotvecB}
},
\\
\constraintCouplingQAQBRot = \constraintCouplingQBQARot = \matrix{
    \tns{0} & \tns{0} \\
    \tns{0} & \tns{0}
},
\quad
\constraintCouplingLQARot = \matrix{
\tns{0} & \pfrac{\rotvecTildeBA}{\rotvecA} \Ttrans\br{\rotvecA}
}
\quad\text{and}\quad
\constraintCouplingLQBRot = \matrix{
\tns{0} & & \pfrac{\rotvecTildeBA}{\rotvecB} \Ttrans\br{\rotvecB}
}
.
\end{aligned}
\end{align}
In practice, all derivatives explicitly stated in~\eqref{eq:jacobian_rotational_coupling} are evaluated using forward automatic differentiation (FAD), cf.~\cite{Korelc2016}, using the Sacado software package~\cite{SacadoWebsite}, which is part of the Trilinos project~\cite{Mayr2026}.

\section{Coupling equations in a discretized setting}
\label{sec:discretized_coupling_equations}

In this section, we illustrate how the coupling equations presented in \Cref{sec:governing_equations} can be incorporated into a discretized setting, i.e., how the kinematic variables and generalized forces are mapped to the discrete degrees of freedom of the underlying beam formulation.
The derivation is demonstrated for the Lagrange multiplier formulation, but can be stated for the penalty regularization in an analogous manner.
We do not assume a specific beam formulation, but present the equations in a general form such that they can be applied to any geometrically exact beam formulation, independent of the choice of primary variables and interpolation schemes.
For clarity, we restrict the following derivation to the interaction of two beam elements only.
The extension to multiple interactions and the assembly into the global system follow standard finite element procedures.
Each beam element is assumed to possess~$\nDofBeami$ discrete degrees of freedom.
The mapping between the discrete variations of the beam element degrees of freedom and the generalized variations of the cross-section kinematics is given by
\begin{align}
\label{eq:discrete_variation_mapping}
\vector{
    \drbeami \\
    \drotmulti
}
=
\Hbeami(\xiBeami,\qbeami) \dqbeami
.
\end{align}
Here,~$\xiBeami \in \R{}$ denotes the cross-section parameter coordinate along the beam axis,~$\qbeami \in \R{\nDofBeami}$ collects the discrete degrees of freedom of a single beam element,~$\dqbeami \in \R{\nDofBeami}$ represents their variations and~$\Hbeami \in \R{6\times \nDofBeami}$ is the shape function matrix for the variations (which in general depends non-linearly on the current deformation state).
It is emphasized that the two beam elements can be of different beam formulation type and also different discretization scheme, i.e.,~$\nDofBeamA \neq \nDofBeamB$.
Using~\eqref{eq:discrete_variation_mapping}, the variational form~\eqref{eq:lagrange_multiplier_weak_form_matrix_vector} of the coupling constraints can be expressed in terms of the discrete degrees of freedom as
\begin{align}
\label{eq:discrete_weak_form_coupling}
\dWLagrange_h =
    \dqbeamA\tr \HbeamA\tr
    \underbrace{
    \br{
        \constraintCouplingQALPos \lagrangePos
        +
        \constraintCouplingQALRot \lagrangeRot
    }}_{\rBAA}
    +
    \dqbeamB\tr \HbeamB\tr
    \underbrace{\br{
        \constraintCouplingQBLPos \lagrangePos
        +
        \constraintCouplingQBLRot \lagrangeRot
    }}_{\rBAB}
    +
    \dlagrangePos\tr \couplingPos
    +
    \dlagrangeRot\tr \couplingRot
.
\end{align}
Here, the subscript~$h$ indicates that the variational form is evaluated in a discretized setting and~$\rBAi \in \R{\nDofBeami}$ denotes the nodal forces and moments corresponding to the coupling forces and moments acting on the cross-sections.
Moreover, the explicit dependence of the shape function matrices on the parameter coordinate and the discrete degrees of freedom has been omitted for brevity.
The mapping of the discrete increments to the increments of the cross-section kinematics is given by
\begin{align}
\label{eq:discrete_increment_mapping}
\vector{
    \Drbeami \\
    \Drotmulti
}
=
\Nbeami(\xiBeami,\qbeami) \Dqbeami
,
\end{align}
where~$\Nbeami \in \R{6\times \nDofBeami}$ denotes the shape function matrix for the increments.
For the consistent linearization of~\eqref{eq:discrete_weak_form_coupling}, we need to account for the linearization of the shape functions as well as the linearization of the coupling forces, i.e.,
\begin{align}
\Delta \br{ \Hbeami\tr \rBAi }
=
\underbrace{\Delta \Hbeami\tr \rBAi}_{\DHbeamLini(\rBAi) \Dqbeami} + \Hbeami\tr \Delta \rBAi
.
\end{align}
Usually, the linearization of the shape functions~$\Delta \Hbeami\tr$ is not evaluated explicitly and applied to~$\rBAi$.
Instead, the expression~$\DHbeamLini(\rBAi) \in \R{\nDofBeami \times \nDofBeami}$ is computed, which maps the discrete increments~$\Dqbeami$ to the stiffness contributions resulting from the linearization of the shape functions.
With this, we can state the linearization of~\eqref{eq:discrete_weak_form_coupling} as
\begin{align}
\begin{aligned}
\Delta \vector{ \rBAA \\ \rBAB \\ \couplingPos \\ \couplingRot }
=
\left(
\matrix{
    \HbeamA\tr & \mat{0} & \mat{0} & \mat{0} \\
    \mat{0} & \HbeamB\tr & \mat{0} & \mat{0} \\
    \mat{0} & \mat{0} & \matI & \mat{0} \\
    \mat{0} & \mat{0} & \mat{0} & \matI
}
\matrix{
    \constraintCouplingQAQAPos + \constraintCouplingQAQARot & \constraintCouplingQAQBPos + \constraintCouplingQAQBRot & \constraintCouplingQALPos & \constraintCouplingQALRot \\
    \constraintCouplingQBQAPos + \constraintCouplingQBQARot & \constraintCouplingQBQBPos + \constraintCouplingQBQBRot & \constraintCouplingQBLPos & \constraintCouplingQBLRot \\
    \constraintCouplingLQAPos & \constraintCouplingLQBPos & \mat{0} & \mat{0} \\
    \constraintCouplingLQARot & \constraintCouplingLQBRot & \mat{0} & \mat{0}
}
\matrix{
    \NbeamA & \mat{0} & \mat{0} & \mat{0} \\
    \mat{0} & \NbeamB & \mat{0} & \mat{0} \\
    \mat{0} & \mat{0} & \matI & \mat{0} \\
    \mat{0} & \mat{0} & \mat{0} & \matI
}
\right.
\\
\left.
+
\matrix{
    \DHbeamLinA(\rBAA) & \mat{0} & \mat{0} & \mat{0} \\
    \mat{0} & \DHbeamLinB(\rBAB) & \mat{0} & \mat{0} \\
    \mat{0} & \mat{0} & \mat{0} & \mat{0} \\
    \mat{0} & \mat{0} & \mat{0} & \mat{0}
}
\right)
\vector{ \DqbeamA \\ \DqbeamB \\ \DlagrangePos \\ \DlagrangeRot }
.
\end{aligned}
\end{align}

\begin{remark}
The parameter coordinates~$\xiBeami$ defining the interacting cross-sections are obtained by a closest-point projection (CPP) between the two beam centerlines in the reference configuration.
This procedure is well-defined as long as the interaction angle between the beam centerlines is sufficiently large.
For the configurations considered in the present work, this condition is satisfied.
For very small interaction angles, a line-to-line coupling formulation would be required, which is beyond the scope of the present work.
For more details on the CPP in the context of point-to-point interactions, the reader is referred to~\cite{Meier2017}.
\end{remark}

\section{Examples}
\label{sec:examples}

In this section, we present several examples to demonstrate the applicability of the proposed coupling formulation.
In \Cref{sec:example_elbow}, we investigate an L-shaped beam structure and show that the presented coupling formulation is able to exactly model the connection between the two beams.
In \Cref{sec:example_two_crossed_beams}, we investigate the spatial convergence behavior of a coupled beam system using the proposed coupling approach.
Furthermore, we demonstrate that both the Lagrange multiplier and the penalty regularization formulations are objective.
In \Cref{sec:example_double_helix}, we present a dynamic simulation of a double-helix structure, demonstrating that the proposed coupling formulation can be used in combination with \sr and \kl beams and with various beam interpolations and rotational parameterizations.
Finally, in \Cref{sec:example_cylinder}, we investigate the buckling behavior of a wire-wound cylinder.

All numerical examples are set up using the open-source beam finite element pre-processor BeamMe~\cite{BeamMe} and are simulated with the open-source multi-physics research code 4C~\cite{FourCWebsite}.
The resulting nonlinear systems of equations are solved using a Newton--Raphson scheme.

\subsection{L-shaped beam structure}
\label{sec:example_elbow}

\begin{figure}
\centering
\subfigure[]{%
\label{fig:examples_elbow_problem_a}%
\includegraphics[scale=1]{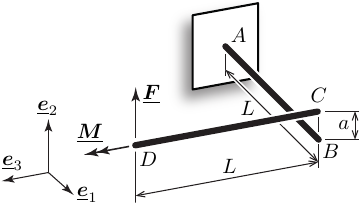}%
}
\hfil
\subfigure[]{%
\label{fig:examples_elbow_problem_b}%
\input{figures/examples_elbow_results.tex}%
}
\caption{L-shaped beam example:~\subref{fig:examples_elbow_problem_a} problem setup and~\subref{fig:examples_elbow_problem_b} deformed structure for an offset~$a=2R$, the simulation is performed with the proposed coupling approach and the Lagrange multiplier method.
For~\subref{fig:examples_elbow_problem_b}, the contour plot visualizes the displacement magnitude in the structure.}
\label{fig:examples_elbow_problem}
\end{figure}

In this first example, we validate the proposed coupling formulation by comparison with alternative approaches for modeling the connection between two beams.
To this end, we consider an L-shaped beam structure as illustrated in \Cref{fig:examples_elbow_problem_a}.
The structure consists of two straight beams of length~$L=\unit[1]{m}$ and circular cross-section with radius~$R=0.05$.
The first beam connects point~$A$ with~$\tns{p}_A = \tnsO$ to point~$B$ with~$\tns{p}_B = L\ex$, whereas the second beam connects point~$C$ with~$\tns{p}_C = L\ex + a\ey$ to point~$D$ with~$\tns{p}_D = L\ex + a\ey + L\ez$.
At point~$A$, the first beam is fully clamped, while at point~$D$ the second beam is loaded with a force~$\tns{F}=\unit[5\cdot 10^{-6}]{N}\ey$ and a moment~$\tns{M}=\unit[5\cdot 10^{-6}]{Nm}\ez$.
At points~$B$ and~$C$, the two beams are connected.
The material parameters of the beams are given by~$\Ebeam=\unit[1]{N/m^2}$ and~$\nubeam = 0$.
Each beam is discretized with~$10$ two-noded linear \sr beam elements, cf.~\cite{Jelenic1999}, and the external loads are applied in~$4$ load steps.

In the first part of this example, we investigate the case of zero offset, i.e.,~$a=0$.
In this case, the points~$B$ and~$C$ coincide in the reference configuration.
This corresponds to a rigid connection between the two beams, i.e., all relative translational and rotational motions at the coupling point are fully constrained.
This allows us to create a reference solution using a direct nodal connection approach, where the nodes at~$B$ and~$C$ share the same degrees of freedom.
We compare this reference solution with the present coupling formulation, where the constraint equations are enforced using both the Lagrange multiplier method and a penalty regularization.
For the penalty regularization, we investigate different penalty parameters~$\penPos = \lambda \Ebeam R$ and~$\penRot = \lambda \Ebeam R^3$, where~$\lambda$ is a dimensionless scaling factor.
In \Cref{tbl:example_elbow_no_offset_results}, the beam positions at point~$D$ are listed for the different coupling formulations and constraint enforcement methods.
The nodal connection approach and the present coupling formulation with the Lagrange multiplier method yield the same result up to the tolerance of the linear solver.
In the case of penalty regularization, the results also converge to the reference solution as the penalty parameters increase.
These results confirm the correctness of the proposed coupling formulation in the case of zero offset.
\begin{table}[h]
\caption{Results of the L-shaped beam problem for an offset~$a=0$.
The beam positions~$\tns{r}_D$ at point~$D$ are listed for the proposed coupling formulation with different constraint enforcement methods and compared to a direct nodal connection approach.
For the penalty regularization, results are shown for different scaling factors of the penalty parameter~$\lambda$.}
\label{tbl:example_elbow_no_offset_results}
\begin{tabular}{lll}
\toprule
coupling formulation & constraint enforcement & $\tns{r}_D$ \\
\midrule
proposed & Penalty regularization,~$\lambda = 1$ & $[0.3894111138, 1.0531406309, 0.5758444335]$ \\
 & Penalty regularization,~$\lambda = 10$ & $[0.3949021070, 1.0495488679, 0.5829477796]$ \\
 & Penalty regularization,~$\lambda = 100$ & $[0.3954580049, 1.0491711848, 0.5836652225]$ \\
 & Penalty regularization,~$\lambda = 1000$ & $[0.3955136632, 1.0491332269, 0.5837370378]$ \\
 & Lagrange multiplier & $[0.3955198482, 1.0491290072, 0.5837450180]$ \\
nodal connection & -- & $[0.3955198482, 1.0491290072, 0.5837450180]$ \\
\botrule
\end{tabular}
\end{table}

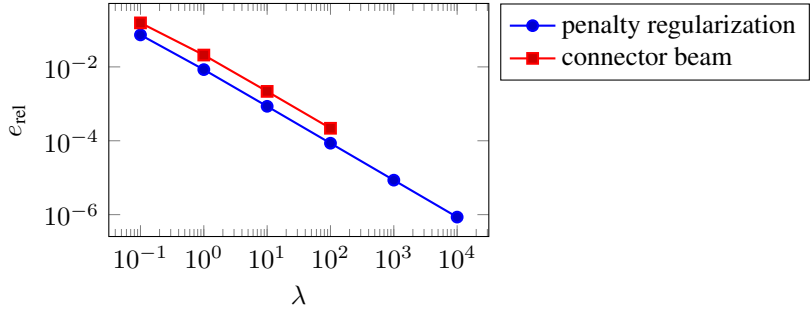
\begin{figure}
\centering
% \tikzsetnextfilename{elbow_convergence}
\begin{tikzpicture}
\begin{axis}[
    xmode=log,
    ymode=log,
    scale only axis, xlabel=$\lambda$, ylabel=$e_\text{rel}$, legend pos=outer north east, legend cell align={left},
    every axis plot post/.append style={
        thick
    },
    ]
\addplot table [col sep=comma, x=penalty, y=error] {figures/examples_elbow_penalty_convergence.csv};
\addlegendentry{penalty regularization}
\addplot table [col sep=comma, x=penalty, y=error] {figures/examples_elbow_connector_convergence.csv};
\addlegendentry{connector beam}
\end{axis}
\end{tikzpicture}
\caption{Relative error of the beam position at point~$D$ with respect to the reference solution obtained with the proposed coupling formulation using Lagrange multipliers for the elbow structure with offset~$a=2R$.
Results are shown for the connector beam approach and the proposed coupling formulation with penalty regularization for different values of the scaling factor~$\lambda$.}
\label{fig:examples_elbow_plot}
\end{figure}

In the second part of this example, we investigate the case of a non-zero offset~$a=2R$.
In this case, an initial offset between the coupling points has to be accounted for in the coupling formulation.
Moreover, the nodal connection approach is not directly applicable.
Therefore, we resort to a different strategy by introducing an additional \emph{connector} beam between points~$B$ and~$C$.
The connector beam is modeled with a single finite element.
For an infinitely stiff connector beam, i.e.,~$E_{c} \to \infty$, this approach would model a rigid connection between the two beams.
Since an infinitely stiff beam cannot be used in practice, the reference solution is obtained with the proposed coupling formulation using Lagrange multipliers.
The reference results are then compared with simulations using connector beams of different stiffness values~$E^{c} = \lambda \Ebeam$.
Additionally, we compare these results with the present coupling formulation using penalty regularization for different penalty parameters~$\penPos = \lambda \Ebeam R$ and~$\penRot = \lambda \Ebeam R^3$.
The deformed structure is shown in \Cref{fig:examples_elbow_problem_b}.
In \Cref{fig:examples_elbow_plot}, the relative error of the beam position at point~$D$, i.e.,~$e_{\text{rel}} = \norm{\tns{r}_{D} - \tns{r}_{D,\text{ref}}}/\norm{\tns{r}_{D,\text{ref}}}$, is shown for the connector beam approach and the penalty regularization for different values of the scaling factor~$\lambda$.
Both approaches converge to the reference solution as~$\lambda$ increases.
With the connector beam approach, the lowest relative error obtained is~$10^{-4}$ for~$\lambda = 10^2$.
Further increasing~$\lambda$ resulted in a non-converging nonlinear solver, likely caused by ill-conditioning of the system due to the high stiffness ratio between the connector beam and the other beams.
For the penalty regularization, the lowest relative error obtained is~$10^{-6}$ for~$\lambda = 10^4$.
As the two approaches are based on fundamentally different modeling strategies, no direct comparison of the errors for equal values of~$\lambda$ should be made.
In summary, the proposed coupling formulation with Lagrange multipliers exactly models a rigid connection between the two beams, also for non-zero offsets, while the penalty regularization provides an accurate approximation.

Finally, we note that comparable Newton--Raphson convergence behavior was observed for the standard nodal coupling and the proposed coupling formulations, with no additional robustness issues associated with the proposed method.

\subsection{Two crossed beams}
\label{sec:example_two_crossed_beams}

\begin{figure}
\centering
\includegraphics[scale=1]{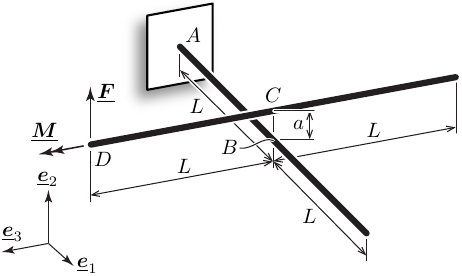}
\caption{Problem setup of the example of two crossed beams.}
\label{fig:examples_two_crossed_beams_problem}
\end{figure}

In this example, we consider two crossed beams as illustrated in \Cref{fig:examples_two_crossed_beams_problem}.
The problem setup is similar to the previous example, the only difference is that instead of an L-shaped structure, we now have two beams with length~$L=\unit[2]{m}$ crossing each other at their midpoint.
The loading, material and load stepping parameters are the same as in the previous example, the offset between the beams is~$a=2R$.
By modifying the problem setup such that the beams cross each other at their midpoint, we can consider general coupling scenarios where the coupling points are located inside the finite element instead of at the nodes.

\begin{figure}
\centering
% \tikzsetnextfilename{two_crossed_beams_convergence}
\begin{tikzpicture}
\begin{axis}[
    xmode=log,
    ymode=log,
    scale only axis, xlabel=$n_e$, ylabel=$e_\text{rel}$ in $\unit{m}$, legend pos=outer north east, legend cell align={left},
    ]
\addplot+ [only marks] table [col sep=comma, x expr=\thisrow{elements}, y=error] {figures/examples_two_crossed_beams_convergence_odd.csv};
\addlegendentry{$n_e$ odd}
\addplot+ [only marks] table [col sep=comma, x expr=\thisrow{elements}, y=error] {figures/examples_two_crossed_beams_convergence_even.csv};
\addlegendentry{$n_e$ even}
\addplot[
    domain=10:500,
    samples=2,
    dashed
] {2*x^(-1)};
\addlegendentry{$\order{h}$}
\addplot[
    domain=3:125,
    samples=2,
] {5e-2*x^(-2)};
\addlegendentry{$\order{h^2}$}
\end{axis}
\end{tikzpicture}
\caption{Convergence plot under spatial mesh refinement of the example of two crossed beams.
The coupling is enforced with the proposed coupling formulation using Lagrange multipliers.
Different convergence behavior can be observed for meshes with even and odd number of elements.
}\label{fig:examples_two_crossed_beams_convergence}
\end{figure}
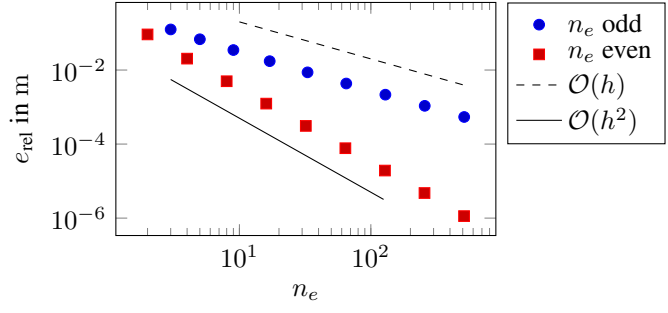

We first investigate the spatial convergence behavior of the proposed coupling formulation for this problem.
The resulting convergence plot for the present coupling formulation with Lagrange multipliers is shown in \Cref{fig:examples_two_crossed_beams_convergence}.
On the ordinate, the relative error of the beam position~$e_{\text{rel}}$ is shown while the abscissa shows the number of finite elements~$n_e$ per beam.
Simulations are performed for meshes with $n_e=2^k$ and $n_e=2^k+1$ with $k=1,\dots,9$.
This results in meshes with even and odd number of elements for each beam.
The reference solution is obtained with a fine discretization of~$n_e = 2048$.
Two-noded \sr beam elements are employed for spatial discretization, cf.~\cite{Jelenic1999}.
For a pure beam problem, the expected convergence order for the beam positions is~$\order{h^2}$, with~$h=2L/n_e$ being the beam finite element length.
One can see that the convergence behavior differs for even and odd number of elements.
For meshes with an even number of elements, the obtained convergence order is the expected optimal order of~$\order{h^2}$.
However, for meshes with an odd number of elements, the convergence order is reduced to~$\order{h}$.
This behavior can be attributed to the fact that for meshes with an even number of elements, the coupling points are located at discrete nodes of the finite element mesh, while for meshes with an odd number of elements, the coupling points are located exactly in the middle of the central finite element of each beam.
In the derivation of the governing weak form of the employed \sr beam formulation, discrete forces and moments only appear at inter-element boundaries~\cite{Jelenic1999,Meier2019}.
When the coupling point coincides with a node (even meshes), the discrete constraint forces and moments directly act at element boundaries (i.e., nodes), preserving the optimal convergence order.
When it lies inside an element (odd meshes), the discrete constraint forces and moments act in the middle of an element, thus degrading the convergence order.
It is important to note that this effect is not caused by the presented coupling formulation itself, but is an inherent property of geometrically exact beam formulations.
We can conclude that coupling points should, if possible, be chosen to coincide with finite element nodes, but even if this is not the case, the formulation still provides a converging solution.

\begin{figure}
\centering
% \tikzsetnextfilename{two_crossed_beams_objectivity_lagrange}
\begin{tikzpicture}
\begin{axis}[
    scale only axis, xlabel=load step, ylabel=$\Pi_\text{int}$ in $\unit{Nm}$, legend pos=outer north east, legend cell align={left},
    every axis plot post/.append style={
        thick, mark=none
    },
    title=Lagrange multipliers
    ]
\addplot table [col sep=comma, x=time, y=internal_energy] {figures/examples_two_crossed_beams_objectivity_lagrange.csv};
\end{axis}
\end{tikzpicture}
\hfil
% \tikzsetnextfilename{two_crossed_beams_objectivity_penalty}
\begin{tikzpicture}
\begin{axis}[
    scale only axis, xlabel=load step, ylabel=$\Pi_\text{int} + \Pi^\epsilon$ in $\unit{Nm}$, legend pos=outer north east, legend cell align={left},
    every axis plot post/.append style={
        thick, mark=none
    },
    title=Penalty regularization
    ]
\addplot table [col sep=comma, x=time, y=total_energy] {figures/examples_two_crossed_beams_objectivity_penalty.csv};
\end{axis}
\end{tikzpicture}
\caption{Evolution of the total elastic energy for the example of two crossed beams with~$n_e = 9$.
The plot on the left shows the Lagrange multiplier approach, while the one on the right shows the penalty regularized approach.
For the penalty case, the total elastic energy includes penalty contributions.
}\label{fig:examples_two_crossed_beams_objectivity}
\end{figure}
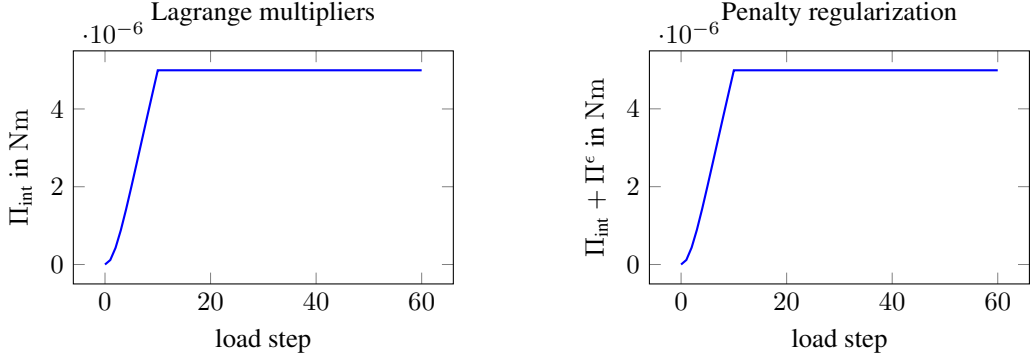

Finally, we investigate the objectivity of the proposed coupling formulation for both the Lagrange multiplier and the penalty regularization approach.
To this end, we take the example of two crossed beams with~$n_e = 9$ and apply the external load in~$10$ load steps.
We then apply a rotation around the~$\ex$-axis with the rotation angle of~$2\pi$ at point~$A$ in a Dirichlet manner.
All external forces are rotated accordingly.
The rotation is applied in~50 additional load steps.
\Cref{fig:examples_two_crossed_beams_objectivity} shows the evolution of the total elastic energy of the system for both the Lagrange multiplier and the penalty regularization approach.
In the Lagrange multiplier case, the total elastic energy only consists of the internal elastic energy of the beams, while in the penalty regularization case, the total elastic energy also includes the contribution of the penalty potential.
It can be seen that for both cases, the energy increases for the first~10 load steps when the external load is applied and then stays constant up to non-linear solver tolerance during the rotation of the structure, confirming the objectivity of both approaches.

\subsection{Double helix}
\label{sec:example_double_helix}

\begin{figure}
\newcommand{\spacehorizontal}{-7pt}
\centering
\subfigure[]{%
\label{fig:examples_helix_problem}%
\input{figures/examples_helix_problem.tex}%
}
\subfigure[]{%
\label{fig:examples_helix_result_50}%
\includegraphics[scale=0.12]{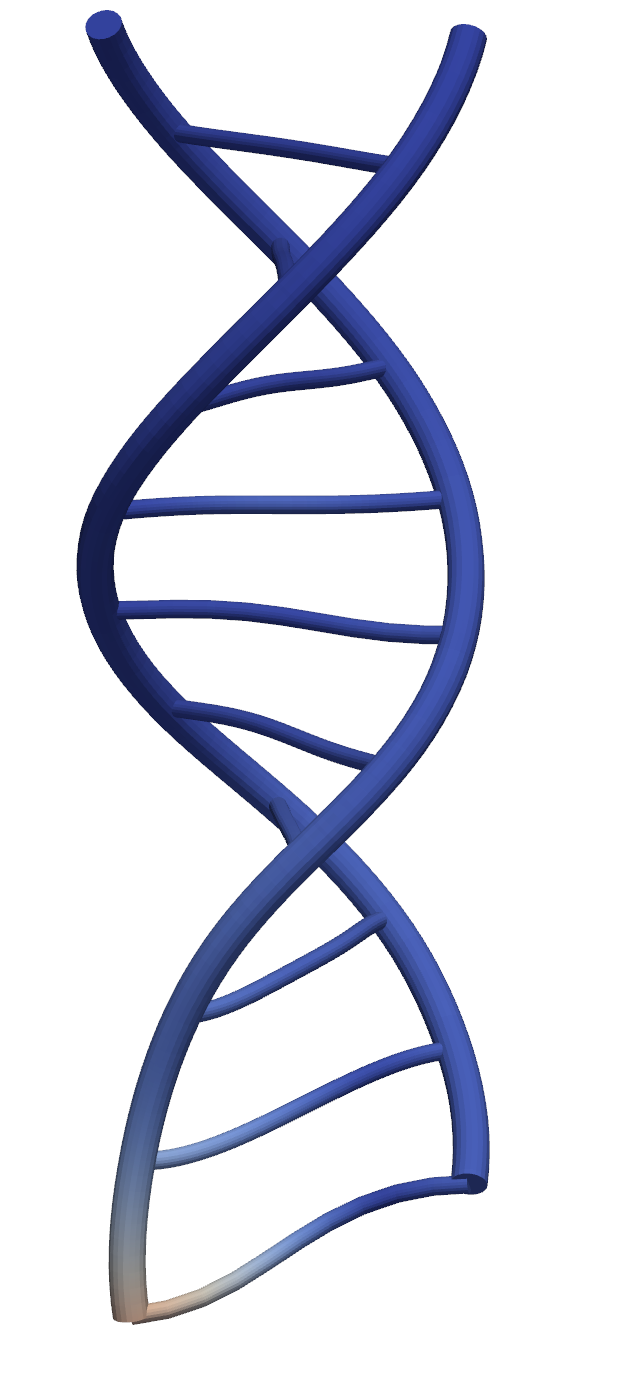}
}
\hspace{\spacehorizontal}
\hspace{\spacehorizontal}
\hspace{\spacehorizontal}
\subfigure[]{%
\label{fig:examples_helix_result_150}%
\includegraphics[scale=0.12]{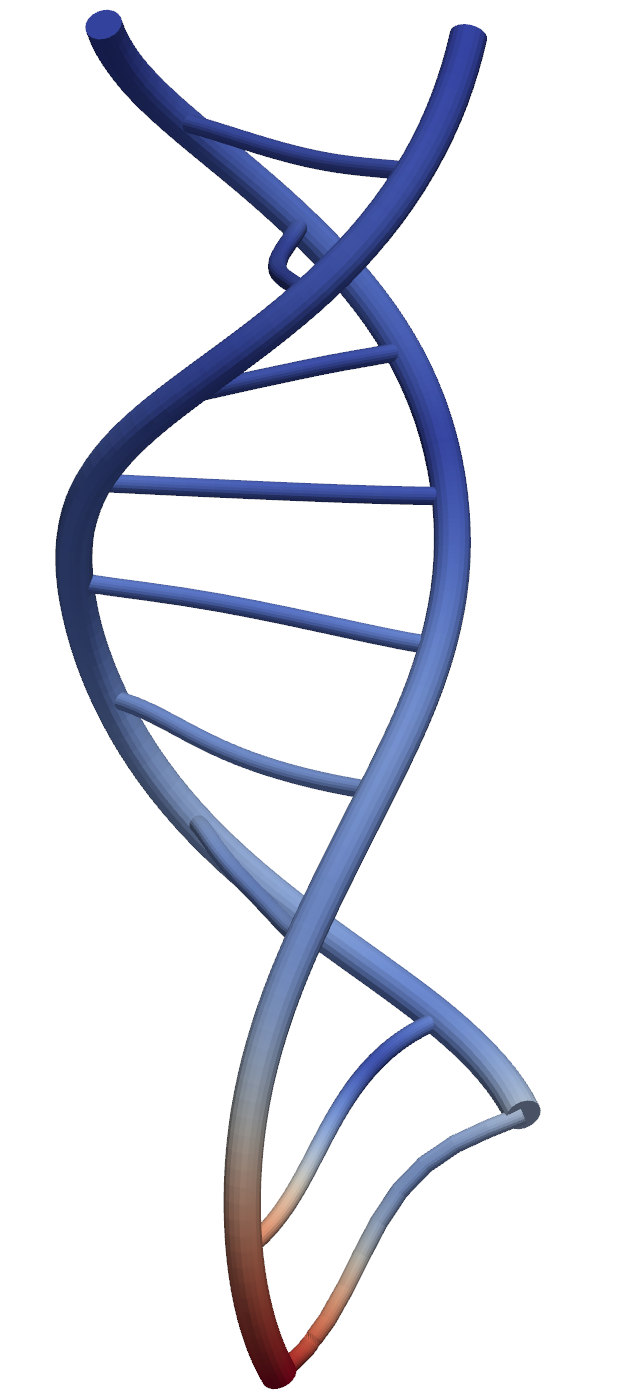}
}
\hspace{\spacehorizontal}
\hspace{\spacehorizontal}
\subfigure[]{%
\label{fig:examples_helix_result_250}%
\includegraphics[scale=0.12]{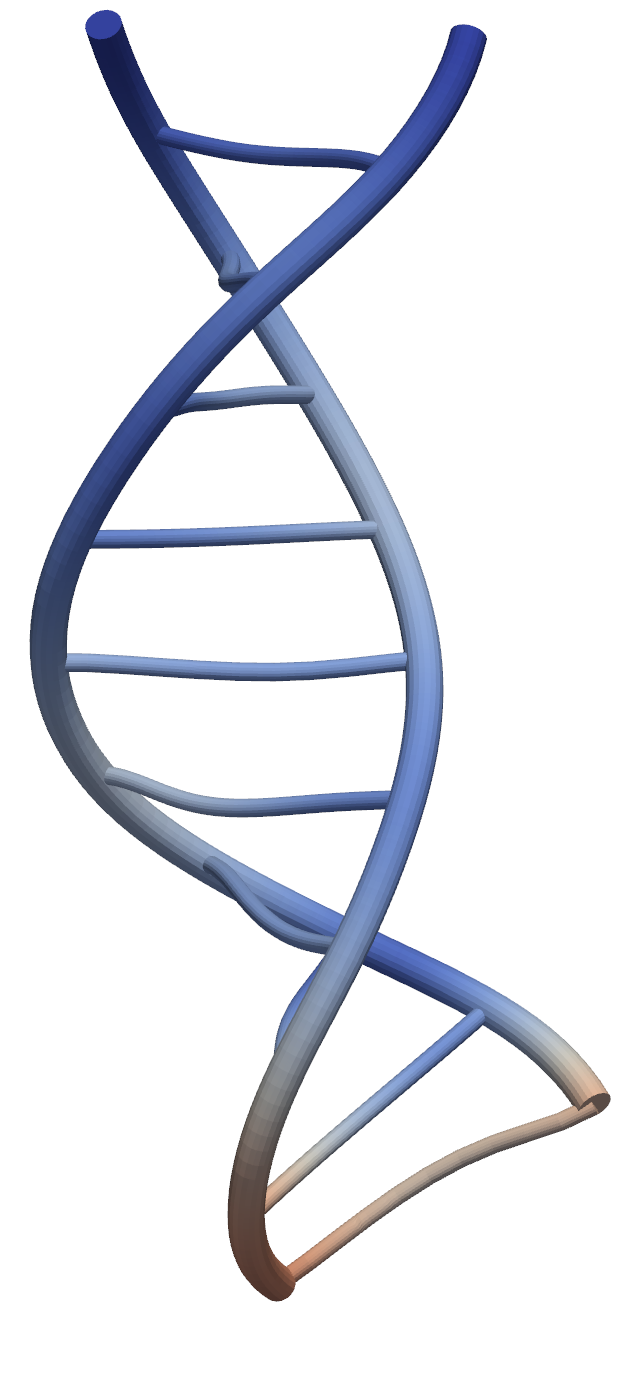}
}
\hspace{\spacehorizontal}
\input{figures/examples_helix_result_color_bar.tex}
\caption{Double helix example:~\subref{fig:examples_helix_problem} problem setup,
\subref{fig:examples_helix_result_50} deformed configuration at~$t = \unit[50]{s}$,
\subref{fig:examples_helix_result_150} deformed configuration at~$t = \unit[150]{s}$,
\subref{fig:examples_helix_result_250} deformed configuration at~$t = \unit[250]{s}$.
For the deformed configurations, the contour plots visualize the displacement magnitude in the structure.}
\label{fig:examples_helix}
\end{figure}

In this example, we investigate the dynamic behavior of a double helix structure as shown in~\Cref{fig:examples_helix_problem}.
This example demonstrates the applicability of the proposed coupling formulation in combination with different beam formulations, interpolation schemes, and rotational parameterizations.
The structure consists of two right-handed helices with a radius of~$r_{\text{helix}} = \unit[2]{m}$, a helix angle of~$\alpha_{\text{helix}} = \pi/4$, and a twist angle of~$\beta_{\text{helix}} = 2\pi$, which corresponds to a helix height of~$h_{\text{helix}} = \unit[12.566]{m}$.
The axis of the helices is aligned with the~$\ez$-axis and the two helices are phase-shifted by~$\pi$.
In between the two helices, there are~$n_{\text{connector}} = 10$ connecting beams.
These connecting beams are placed in the~$\ex$--$\ey$ plane and are equally spaced along the helix height.

The beams forming the helices have a circular cross-section with a radius of~$R_{\text{helix}} = \unit[0.2]{m}$, while the connecting beams have a circular cross-section with a radius of~$R_{\text{connector}} = \unit[0.1]{m}$.
The material parameters of all beams are given by~$\Ebeam=\unit[1]{N/m^2}$,~$\nubeam = 0$, and~$\density = \unit[1]{kg/m^3}$.
At the top of the structure, the points~$A$ and~$B$ are fully clamped.
At the bottom of the structure, point~$C$ is subjected to a downward acting force~$\tns{F}(t)$, while no boundary conditions are applied at point~$D$.
The force is linearly increased from~$t=\unit[0]{s}$ to~$t=\unit[20]{s}$, reaching a value of~$\tns{F} = -\unit[10^{-3}]{N}\ez$, and then linearly decreased back to zero until~$t=\unit[40]{s}$.

In \Cref{fig:examples_helix_problem}, the color coding indicates the beam finite element formulations employed for the individual beams, which are listed in \Cref{tbl:example_helix_fem_formulations}.
The two helices are modeled using \sr beam elements with a 3rd-order Hermite interpolation of the beam centerline and a 2nd-order Lagrangian interpolation of the relative rotation vectors.
The connecting beams alternate between \sr beams with 1st- to 4th-order Lagrangian interpolation and \kl beam elements with a 3rd-order Hermite interpolation of the beam centerline and a 2nd-order Lagrangian interpolation of the twist.
All \sr beams parametrize the cross-section orientation using rotation vectors, whereas the \kl beams use the (normalized) tangent vector and a scalar twist.
The choice of different beam formulations is not motivated by physical considerations but rather serves to demonstrate the applicability of the proposed coupling formulation in combination with different beam formulations and interpolation schemes.
The coupling is realized with the proposed coupling formulation and a penalty regularization with penalty parameters~$\penPos = \EbeamAvg \radiusAvg$ and~$\penRot = \EbeamAvg \radiusAvg^3$.
Each connector beam is discretized with~10 finite elements, while the helix starting at point~$A$ consists of~17 elements and the one starting at point~$B$ consists of~19 elements.
This discretization leads to general coupling scenarios in which the coupling points are located at arbitrary positions within the finite elements.
The simulation is performed until~$t = \unit[250]{s}$ with a constant time step size of~$\Delta t = \unit[1]{s}$ using a generalized-$\alpha$ Lie group time integration method~\cite{Bruels2010,Bruels2012}.
The parameters of the time integration scheme are chosen as $\alpha_m = 0.5$, $\alpha_f = 0.5$, $\beta = 0.25$, and $\gamma = 0.5$, corresponding to a non-dissipative scheme with spectral radius $\rho_\infty = 1$.
\begin{table}
\caption{Different beam finite element formulations employed in the double helix example.
The typeID refers to the color coding in \Cref{fig:examples_helix_problem}.}\label{tbl:example_helix_fem_formulations}%
\begin{tabular}{llllll}
\toprule
type ID & formulation  & interpolation pos. & interpolation rot. & rot. parametrization & \cf \\
\midrule
1    & \sr   & 3rd-order Hermite  & 2nd-order Lagrange & rotation vector & \cite{Meier2019} \\
2    & \sr   & 1st-order Lagrange  & 1st-order Lagrange & rotation vector & \cite{Jelenic1999} \\
3    & \sr   & 2nd-order Lagrange  & 2nd-order Lagrange & rotation vector & \cite{Jelenic1999} \\
4    & \sr   & 3rd-order Lagrange  & 3rd-order Lagrange & rotation vector & \cite{Jelenic1999} \\
5    & \sr   & 4th-order Lagrange  & 4th-order Lagrange & rotation vector & \cite{Jelenic1999} \\
6    & \kl   & 3rd-order Hermite & 2nd-order Lagrange & tangent + twist  & \cite{Meier2015,Meier2014} \\
\botrule
\end{tabular}
\end{table}

\begin{figure}
\centering
% \tikzsetnextfilename{helix_time_displacement}
\begin{tikzpicture}
\begin{axis}[
    scale only axis, xlabel=$t$ in $\unit{s}$, ylabel=$u$ in $\unit{m}$, legend pos=outer north east, legend cell align={left},
    every axis plot post/.append style={
        thick,
        mark=none
    },
    ]
\addplot table [col sep=comma, x=time, y=displacement_x] {figures/examples_helix_time_displacement.csv};
\addlegendentry{$u_1$}
\addplot table [col sep=comma, x=time, y=displacement_y] {figures/examples_helix_time_displacement.csv};
\addlegendentry{$u_2$}
\addplot table [col sep=comma, x=time, y=displacement_z] {figures/examples_helix_time_displacement.csv};
\addlegendentry{$u_3$}
\end{axis}
\end{tikzpicture}
\caption{Time evolution of the helix displacements at the point of action of the force.}\label{fig:examples_helix_time_displacement}
\end{figure}
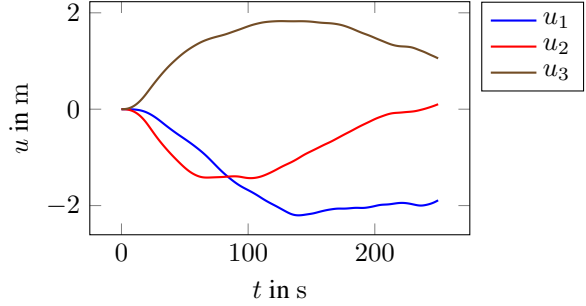

\Cref{fig:examples_helix_result_50,fig:examples_helix_result_150,fig:examples_helix_result_250} show the deformed structure at different times during the simulation, while \Cref{fig:examples_helix_time_displacement} shows the time evolution of the displacement at the point of action of the force, i.e., point~$C$.
It can be observed that the structure undergoes large deformations and exhibits complex dynamic behavior.
Moreover, this example demonstrates that the proposed coupling scheme is capable of handling different beam formulations and interpolation schemes in a unified manner.

\subsection{Wire-wound cylinder}
\label{sec:example_cylinder}

\begin{figure}
\centering
\includegraphics[scale=1]{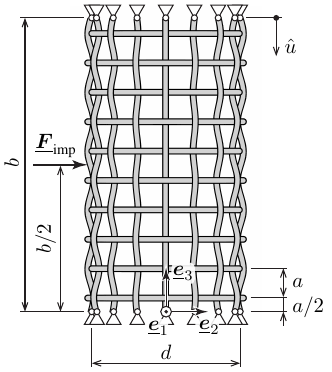}
\caption{Wire-wound cylinder problem setup.}\label{fig:examples_cylinder_problem}
\end{figure}

In this final example, we investigate the buckling behavior of a wire-wound cylinder as shown in \Cref{fig:examples_cylinder_problem}.
The cylinder has a diameter of~$d = \unit[2]{m}$ and its axis is aligned with the $\ez$-axis.
The structure is formed by two orthogonal families of slender fibers.
One family consists of~$n_{\text{circ}}=10$ circumferential rings which are equally spaced along the cylinder axis, while the other consists of~$n_{\text{axi}}=16$ axial fibers which are equally spaced in the circumferential direction.
The distance between the circumferential rings is the same as the distance between the axial fibers (measured along the circumferential arc length) and amounts to~$a = \pi d / n_{\text{axi}}$.
The total height of the structure is~$b = a \, n_{\text{circ}}$.
Each fiber has a circular cross-section with a radius of~$R=\unit[0.04]{m}$.
The fibers are interwoven with each other such that the beam cross-sections exactly touch at the intersection points.
The reference position of the axial fibers can be given in parametric form:
\begin{align}
\rbeam_{\text{axi},0}^i (z) = \br{\frac{d}{2} - \br{-1}^{i} R \sin \br{\pi \frac{z}{a}}}
\br{\cos\br{i \frac{2\pi}{n_{\text{axi}}}} \ex + \sin\br{i \frac{2\pi}{n_{\text{axi}}}} \ey}
+ z \ez,
\end{align}
with $i=1,\dots,n_{\text{axi}}$ and $z \in [0,b]$.
The reference position of the circumferential fibers can be given in parametric form:
\begin{align}
\rbeam_{\text{circ},0}^i (\phi) = \br{\frac{d}{2} - \br{-1}^{i} R \cos\br{\phi \frac{n_{\text{axi}}}{2}}}
\br{\cos\phi \ex + \sin\phi \ey}
+ a \br{i - \frac{1}{2}} \ez,
\end{align}
with $i=1,\dots,n_{\text{circ}}$ and $\phi \in [0,2\pi)$.

The fibers material parameters are given by~$\Ebeam=\unit[1]{N/m^2}$ and~$\nubeam = 0$.
The start and end points of the axial fibers are completely clamped, for the end points (at the top of the cylinder) a displacement~$\hat{u}=\unit[0.2]{m}$ is prescribed in negative~$\ez$-direction.
To avoid numerical issues and trigger the overall buckling behavior, a small imperfection load~$\tns{F}_{\text{imp}}=\unit[10^{-6}]{N} \,\ey$ is applied to the structure at the point~$\rbeamO = -d/2 \, \ex + b/2 \, \ez$.
No other external loads are applied to the structure.

The circumferential and axial fibers are assumed to be rigidly connected at the~160 intersection points.
These connections are modeled with the proposed beam-to-beam point couplings and a Lagrange multiplier constraint enforcement.
Geometrically exact \sr beam finite elements with a~$C^1$-continuous third order Hermite interpolation of the centerline are used for the spatial discretization.
Each circumferential fiber is discretized with~24 beam finite elements, while each axial fiber is discretized with~15 beam finite elements.
The prescribed displacement~$\hat{u}$ is applied in~100 quasi-static load steps.
The imperfection load is applied in the first load step and kept constant in all subsequent load steps.

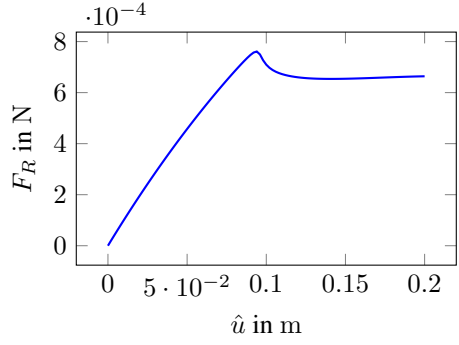
\begin{figure}
\centering
% \tikzsetnextfilename{cylinder_force_displacement}
\begin{tikzpicture}
\begin{axis}[
    scale only axis, xlabel=$\hat{u}$ in $\unit{m}$, ylabel=$F_R$ in $\unit{N}$, legend pos=outer north east, legend cell align={left},
    every axis plot post/.append style={
        thick,
        mark=none
    },
    ]
\addplot table [col sep=comma, x=displacement, y=force] {figures/examples_cylinder_force_displacement.csv};
\end{axis}
\end{tikzpicture}
\caption{Force-displacement curve for the wire-wound cylinder.}\label{fig:examples_cylinder_force_displacement}
\end{figure}

\begin{figure}
\centering
\newcommand{\factor}{0.23}
\subfigure[]{%
    \label{fig:examples_cylinder_force_deformed_1}%
    \includegraphics[width=\factor\textwidth]{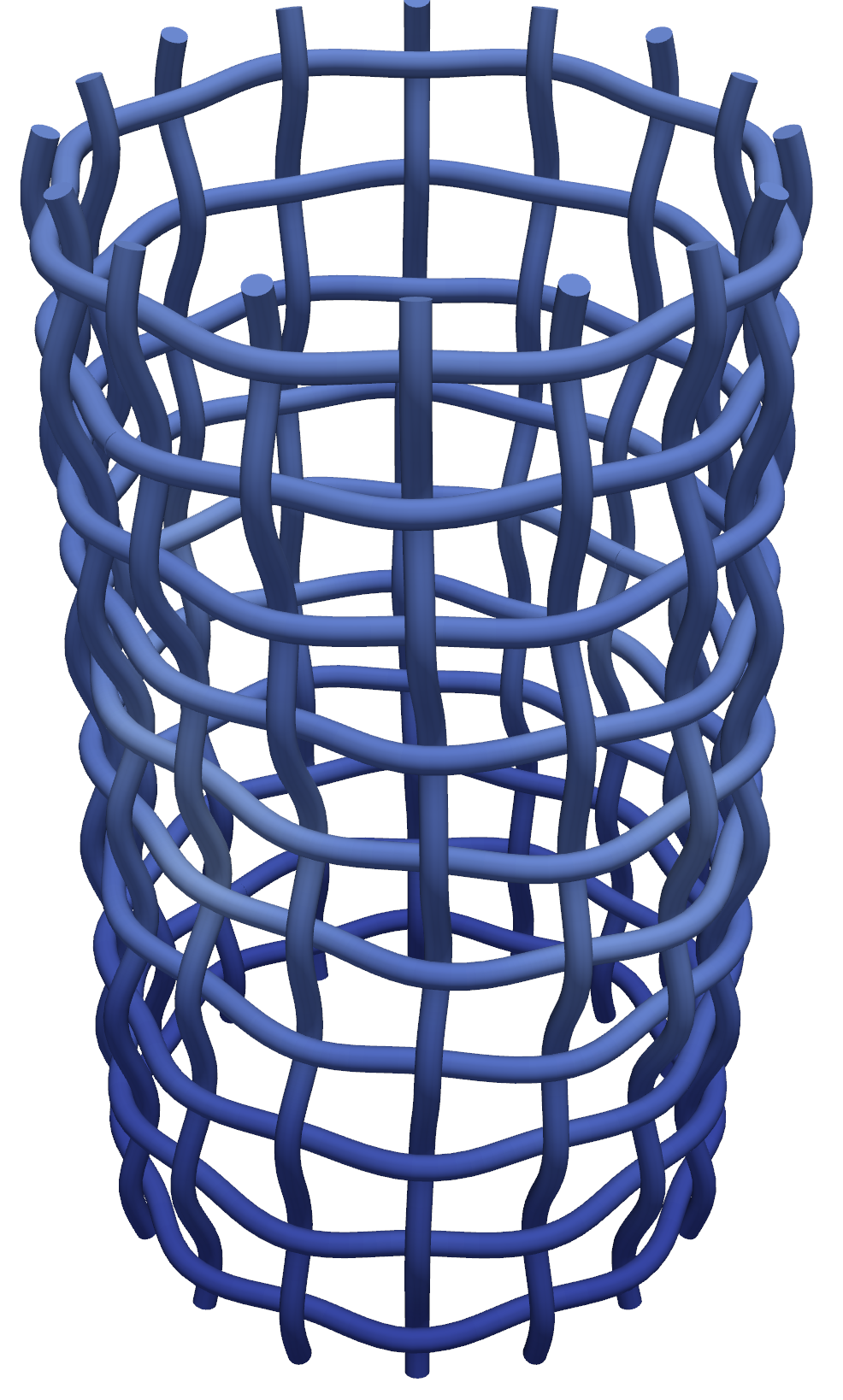}
}
\hfill
\subfigure[]{%
    \label{fig:examples_cylinder_force_deformed_2}%
    \includegraphics[width=\factor\textwidth]{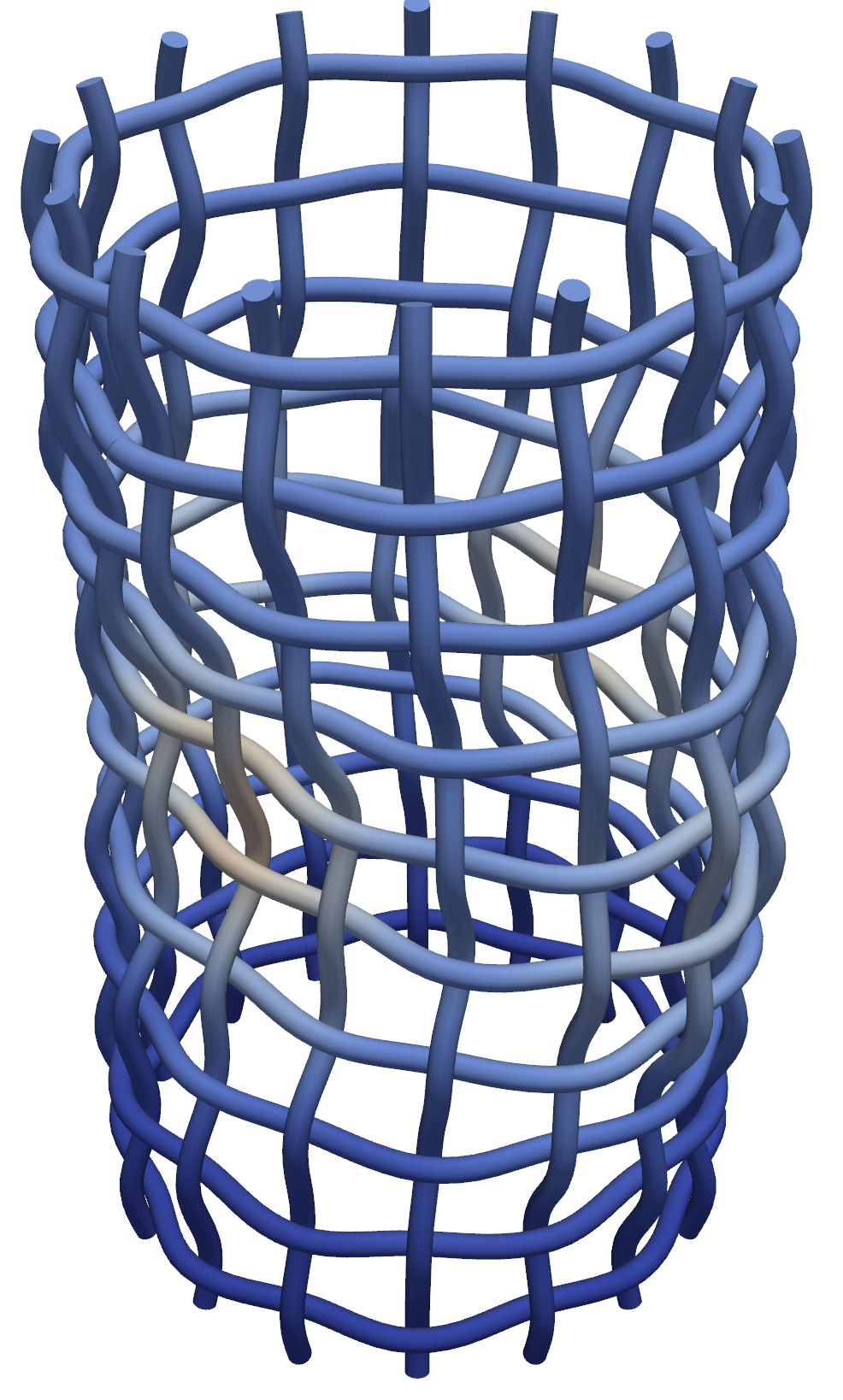}
}
\hfill
\subfigure[]{%
    \label{fig:examples_cylinder_force_deformed_3}%
    \includegraphics[width=\factor\textwidth]{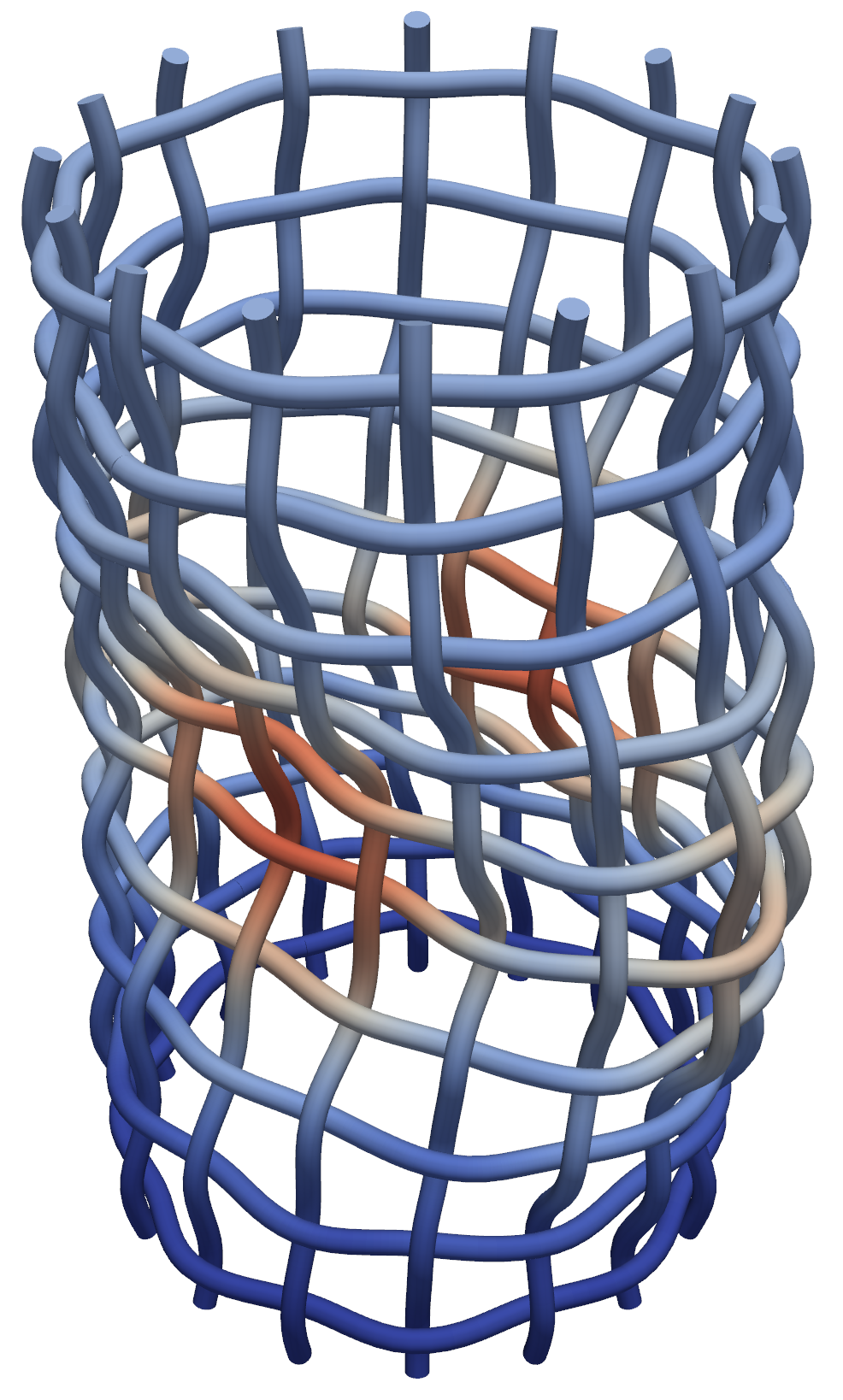}
}
\hfill
\subfigure[]{%
    \label{fig:examples_cylinder_force_deformed_4}%
    \includegraphics[width=\factor\textwidth]{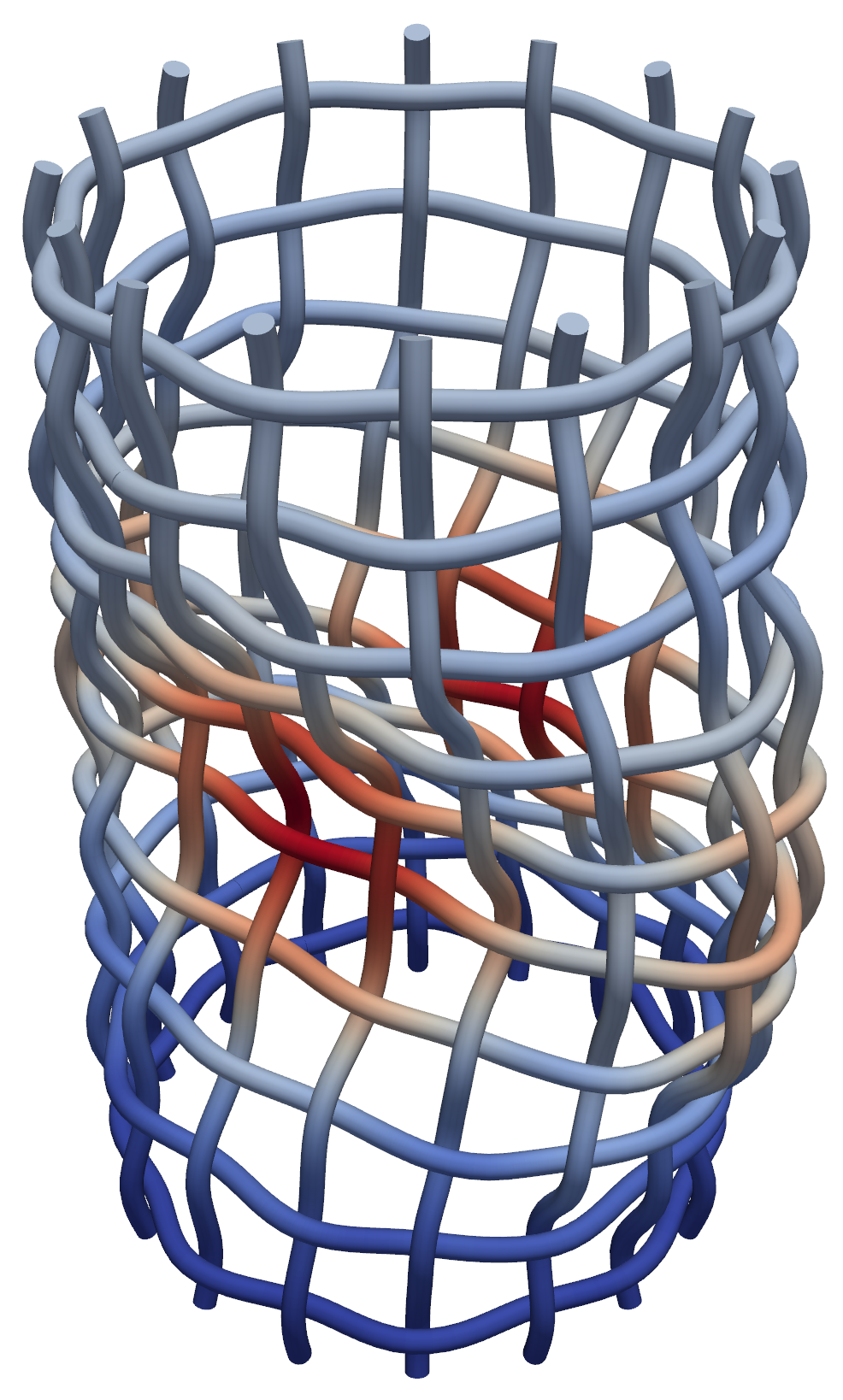}
}
\input{figures/examples_cylinder_result_color_bar.tex}
\caption{Deformed configuration of the wire-wound cylinder at different load steps.
The snapshots are taken at \subref{fig:examples_cylinder_force_deformed_1} $\hat{u} = \unit[0.096]{m}$, \subref{fig:examples_cylinder_force_deformed_2} $\hat{u} = \unit[0.11]{m}$, \subref{fig:examples_cylinder_force_deformed_3} $\hat{u} = \unit[0.16]{m}$ and \subref{fig:examples_cylinder_force_deformed_4} $\hat{u} = \unit[0.2]{m}$.
The contour plots visualize the displacement magnitude in the structure.}\label{fig:examples_cylinder_force_deformed}
\end{figure}

The resulting force-displacement curve is shown in~\Cref{fig:examples_cylinder_force_displacement}, where the reaction force~$F_R$ is plotted against the applied displacement~$\hat{u}$.
The force~$F_R$ is the~$\ez$-component of the summed up boundary reaction force at the top of the cylinder.
One can see that up to a displacement of~$\hat{u} \approx \unit[0.09]{m}$, the structure behaves more or less linear.
From that point on the buckling behavior starts to develop.
At~$\hat{u} \approx \unit[0.094]{m}$, the maximum reaction force of~$F_R \approx \unit[7.6\cdot 10^{-4}]{N}$ is reached, followed by a drop off to a roughly constant reaction force of~$F_R \approx \unit[6.5\cdot 10^{-4}]{N}$ for further increasing displacements.
\Cref{fig:examples_cylinder_force_deformed} shows the deformed structure at different representative load steps.
It can be seen that even though the imperfection force is applied in a non-symmetric manner, the overall buckling behavior is symmetric.

This example showcases that the proposed beam-to-beam coupling method can be used to model complex structures with a large number of beam-to-beam connections in an accurate and robust manner.

\section{Conclusion}

In this work, a beam-to-beam point coupling formulation has been presented that enables the modeling of general rigid connections between beam cross-sections.
The proposed approach expresses the coupling constraints solely in terms of the cross-section kinematics of the geometrically exact beam formulation, i.e., centroid positions and orientations, which makes the formulation independent of the specific beam theory and discretization scheme.
As a consequence, beam elements with different interpolation schemes, primary variables, and rotational parameterizations can be coupled in a unified manner.
Constraint enforcement was formulated using the Lagrange multiplier method, resulting in a saddle-point system.
Additionally, a penalty regularization was presented that allows for the condensation of the Lagrange multipliers.
The formulation was analyzed with respect to fundamental properties such as symmetry, objectivity, and consistency with a stress-free reference configuration.
Furthermore, the discrete coupling terms were described in a formulation-independent manner, allowing straightforward integration into existing beam finite element frameworks.
Several numerical examples were presented to demonstrate the accuracy and robustness of the proposed coupling formulation, as well as its applicability to challenging and practically relevant interaction scenarios.

Future work may extend the formulation towards more general interaction scenarios, such as line-to-line beam coupling.

\backmatter

\bmhead{Acknowledgements}

Sketches in this work have been created using the Adobe Illustrator plug-in LaTeX2AI (\url{https://github.com/isteinbrecher/latex2ai}).

\begin{appendices}

\section{Useful identities for large rotations}

In the following, we list useful identities for large rotations that are employed in the analysis of the proposed positional and rotational coupling formulations, cf.~\cite{Meier2023}:
\begin{align}
\label{eq:tangential_map_transformed}
\triad \, \Ttrans \br[0]{\rotvec} \triad\tr
&= \Ttrans \br[0]{\triad \,\rotvec}
\\
\label{eq:skew_map_transformed}
\triad \, \Sskew{\rotvec} \triad\tr
&= \Sskew{\triad \,\rotvec}
\\
\label{eq:log_map_inverse_identity}
\rv \br[0]{\triad} &= -\rv \br[0]{\triad\tr}
\\
\label{eq:log_map_transformed}
\rv \br[0]{\tnss{Q} \triad \tnss{Q}\tr}
&= \tnss{Q} \rv \br[0]{\triad} \quad \forall \, \tnss{Q} \in \SO
.
\end{align}

\section{Properties of the positional coupling formulation}
\label{sec:appendix_properties_positional_coupling}

In the following, we analyze the properties of the deformation measure~$\deformationPos$ and the corresponding variational form~$\dWPos$, thereby confirming that the properties stated in \Cref{sec:requirements_coupling_constraints} are satisfied.

To show that the positional coupling constraint vanishes in the reference configuration (cf.~\eqref{item:stress_free}), we insert the reference positions and orientations into the definition of the positional deformation measure~\eqref{eq:spatial_deformation_measure_pos}, yielding
\begin{align}
\deformationPos
&=
\underbrace{\rbeamBO - \rbeamAO}_{\rbeamBAO} -  \frac{1}{2}\br[1]{\triadAO \underbrace{\RbeamBAOA}_{\br[0]{\triadAO}\tr \rbeamBAO} + \triadBO \underbrace{\RbeamBAOB}_{\br[0]{\triadBO}\tr \rbeamBAO}}
\\
&=
\rbeamBAO -  \frac{1}{2}\br[1]{\rbeamBAO + \rbeamBAO} = \tnsO
.
\end{align}

Next, we show that the deformation measure satisfies the symmetry property~\eqref{item:symmetry}.
To this end, we switch the indices~$1$ and~$2$ in the definition of the positional deformation measure~\eqref{eq:spatial_deformation_measure_pos}, which yields
\begin{align}
\label{eq:symmetry}
\deformationPosSym
&= \rbeamA - \rbeamB - \frac{1}{2}\br{ \triadB \RbeamABOA + \triadA \RbeamABOB }
\\
&= - \rbeamBA + \frac{1}{2}\br{ \triadB \RbeamBAOA + \triadA \RbeamBAOB }
= - \deformationPos
.
\end{align}

To demonstrate objectivity~\eqref{item:objectivity} of the positional coupling formulation, we insert the transformed positions and orientations under a superimposed rigid body motion~\eqref{eq:rigid_body_transformations} into the variational form of the positional coupling constraint~\eqref{eq:weak_form_pos}, i.e.,
\begin{align}
\dWPosStar 
=&\ \br[0]{\dlagrangePos^\ast}\tr \br{
    \rbeamB^\ast - \rbeamA^\ast
    -  \frac{1}{2}\br{ \triadA^\ast \RbeamBAOA + \triadB^\ast \RbeamBAOB }
}
\notag \\
&+
\br[0]{\lagrangePos^\ast}\tr \br{
    \drbeamB^\ast - \drbeamA^\ast
    - \frac{1}{2} \Sskew{\drotmultB^\ast + \drotmultA^\ast}
    \br{\rbeamB^\ast - \rbeamA^\ast} 
}
\\
=&\ \dlagrangePos\tr \br[0]{\rigidBodyRotation}\tr \br{
    \rigidBodyRotation \rbeamB + \rigidBodyTranslation
    - \rigidBodyRotation \rbeamA - \rigidBodyTranslation
    -  \frac{1}{2}\br{ \rigidBodyRotation \triadA \RbeamBAOA
    + \rigidBodyRotation \triadB \RbeamBAOB }
}
\notag \\
\label{eq:objectivity_positional_coupling_intermediate_1}
&+
\lagrangePos\tr \br[0]{\rigidBodyRotation}\tr  \br{
    \rigidBodyRotation \drbeamB - \rigidBodyRotation \drbeamA
    - \frac{1}{2} \Sskew{\rigidBodyRotation \br{\drotmultB + \drotmultA}} \rigidBodyRotation \br{\rbeamB - \rbeamA} 
}
\\
=&\ \dlagrangePos\tr \br[0]{\rigidBodyRotation}\tr \rigidBodyRotation \br{
     \rbeamB - \rbeamA
     -  \frac{1}{2}\br{ \triadA \RbeamBAOA + \triadB \RbeamBAOB }
}
\notag \\
\label{eq:objectivity_positional_coupling_intermediate_2}
&+
\lagrangePos\tr \br[0]{\rigidBodyRotation}\tr \rigidBodyRotation \br{
    \drbeamB - \drbeamA
    - \frac{1}{2} \Sskew{\drotmultB + \drotmultA} \br[0]{\rigidBodyRotation}\tr \rigidBodyRotation \br{\rbeamB - \rbeamA} 
}
\\
\label{eq:objectivity_positional_coupling_intermediate_3}
=&\ \dWPos
.
\end{align}
Between~\eqref{eq:objectivity_positional_coupling_intermediate_1} and~\eqref{eq:objectivity_positional_coupling_intermediate_2}, the identity~\eqref{eq:skew_map_transformed} has been used.
Between~\eqref{eq:objectivity_positional_coupling_intermediate_2} and~\eqref{eq:objectivity_positional_coupling_intermediate_3}, the identity~$\triad\tr \triad = \tnssI$ has been applied.

\section{Properties of the rotational coupling formulation}
\label{sec:appendix_properties_rotational_coupling}

In the following, we analyze the properties of the rotational deformation measure~$\triadTildeBA$ and the corresponding variational form~$\dWRot$, thereby confirming that the properties stated in \Cref{sec:requirements_coupling_constraints} are satisfied.

To show that the rotational coupling constraint vanishes in the reference configuration (cf.~\eqref{item:stress_free}), we insert the reference rotations into~\eqref{eq:relative_rotation_effective}, yielding
\begin{align}
\triadTildeBA
=
\triadBO \br{\triadBO}\tr \triadAO \br{\triadAO}\tr
= \tnssI
,
\end{align}
which directly implies~$\rotvecTildeBA = \tnsO$.

To show that the rotational deformation measure satisfies the symmetry property~\eqref{item:symmetry}, we switch the indices~$1$ and~$2$ in the definition of the relative rotation~\eqref{eq:relative_rotation_effective}, which yields
\begin{align}
\triadTildeAB
=
\triadA \br{\triadAO}\tr \triadBO \br{\triadB}\tr
= (\triadTildeBA)\tr
.
\end{align}
Inserting this into~\eqref{eq:log_map_inverse_identity} yields~$\rotvecTildeBA = -\rotvecTildeAB$.

Before demonstrating objectivity of the variational form, we first analyze the transformation behavior of the relative rotation vector under a superimposed rigid body motion.
Inserting~\eqref{eq:rigid_body_transformations} and~\eqref{eq:relative_rotation_effective} into~\eqref{eq:relative_rotation_vector} yields
\begin{align}
\rotvecTildeBA^\ast
&=
\rv\br[0]{\triadTildeBA^\ast} \\
&=
\rv\br[0]{\triadB^\ast \br{\triadBO}\tr \triadAO \br{\triadA^\ast}\tr} \\
\label{eq:relative_rotation_vector_transformation_intermediate_1}
&=
\rv\br[0]{\rigidBodyRotation \triadB \br{\triadBO}\tr \triadAO \triadA\tr \br{\rigidBodyRotation}\tr} \\
\label{eq:relative_rotation_vector_transformation_intermediate_2}
&=
\rigidBodyRotation \rv\br[0]{\triadB \br{\triadBO}\tr \triadAO \triadA\tr} \\
\label{eq:relative_rotation_vector_transformation}
&=
\rigidBodyRotation \rotvecTildeBA.
\end{align}
Between~\eqref{eq:relative_rotation_vector_transformation_intermediate_1} and~\eqref{eq:relative_rotation_vector_transformation_intermediate_2}, the identity~\eqref{eq:log_map_transformed} has been used.

To demonstrate objectivity~\eqref{item:objectivity} of the variational form of the rotational coupling constraints, we insert~\eqref{eq:rigid_body_transformations} and~\eqref{eq:relative_rotation_vector_transformation} into~\eqref{eq:weak_form_rot}, yielding
\begin{align}
\dWRotStar
&=
\br[0]{\dlagrangeRot^\ast}\tr \rotvecTildeBA^\ast
+
\br[0]{\lagrangeRot^\ast}\tr
\Ttrans\br[0]{\rotvecTildeBA^\ast}
\br{\drotmultB^\ast - \drotmultA^\ast}
\\
\label{eq:variational_form_rotational_coupling_objectivity_intermediate}
&=
\dlagrangeRot\tr \underbrace{\br[0]{\rigidBodyRotation}\tr \rigidBodyRotation}_{=\tnssI} \rotvecTildeBA
+
\lagrangeRot\tr
\underbrace{
    \br[0]{\rigidBodyRotation}\tr
    \Ttrans\br[0]{\rigidBodyRotation\rotvecTildeBA}
    \rigidBodyRotation
}_{=\Ttrans\br[0]{\br[0]{\rigidBodyRotation}\tr \rigidBodyRotation\rotvecTildeBA} = \Ttrans\br[0]{\rotvecTildeBA}}
\br{\drotmultB - \drotmultA} \\
&= \dWRot
.
\end{align}
In~\eqref{eq:variational_form_rotational_coupling_objectivity_intermediate}, the identity~\eqref{eq:tangential_map_transformed} has been used.
Thus, the variational form is invariant under superimposed rigid body motions, confirming objectivity of the rotational coupling formulation.




\end{appendices}


\bibliography{literature_filtered,steinbrecher_filtered,sn-bibliography}% common bib file

\end{document}